\documentclass{aa}
\usepackage{graphics}
\begin{document}

\title{The BeppoSAX view of bright Compton-thin Seyfert 2 galaxies
}
\author{G. Risaliti\inst{1,2}
}

\institute{
Osservatorio Astrofisico di Arcetri, Largo E. Fermi 5, I-50125 Firenze, Italy
\and
Harvard-Smithsonian Center for Astrophysics, 60 Garden Street,
Cambridge, MA, 02138, USA
}
\authorrunning{G. Risaliti}
\titlerunning{BeppoSAX observations of Seyfert 2s}
\offprints{G. Risaliti}

\date{Received / Accepted}

\abstract{
We present the analysis of 31 observations (17 of which are published here for
the first time) of 20 bright Compton thin
Seyfert 2s,
in the 0.1-200 keV band, performed with the BeppoSAX satellite. The sample consists of all
Seyfert 2s in the BeppoSAX public archive, with a 2-10 keV flux
higher than 5$\times10^{-12}$ erg cm$^{-2}$s$^{-1}$.
The good statistics available and the broad energy band permit a detailed study of the main continuum
components of these sources, i.e. the primary power-law, the reflected component, the soft emission and
the high-energy cut-off. The main results of our analysis are: (1) the 3-200 keV intrinsic power-law has
a mean photon index $\Gamma = 1.79 \pm 0.01$, with a dispersion of $\sigma=0.23$. (2) The high-energy
exponential cut-off at E$\sim 100-300$ keV is not an ubiquitous property of Seyfert galaxies: in $\sim 30$\%
of the objects the continuum power-law does not drop up to energies of 300 keV or more. (3) A reflected
component is present in almost all the sources (17 out of 21). The small variations of this component with
respect to the intrinsic continuum, in objects with multiple observations, suggests that the reflector is
not the accretion disk, but must be located much farther from the nucleus. (4) The range of
ratios between the reflected and
intrinsic components suggests that the circumnuclear medium is not homogeneous, and a significant
fraction of the solid angle is covered by a gas thicker than that along the line of sight. (5) The iron K$\alpha$
line is present is all but one the sources. The equivalent width is in the typical range of Seyfert 1s
(EW=100-300 eV) in sources with low absorption (N$_{\rm H} < 3\times 10^{23}$ cm$^{-2}$), and increases
in more absorbed objects, as expected according to unified models. (6) The energy resolution of BeppoSAX
is in general too low to measure the iron line width. However, in 6 cases we measured a significant 
line broadening.
} 
\maketitle



\section{Introduction}
In this paper we present the analysis of the observations of bright Seyfert 2 galaxies performed with
the BeppoSAX satellite.
The main aim of this work is to investigate the average X-ray continuum properties of these sources, through
a homogeneous analysis of all the BeppoSAX archive data. The broad energy band of BeppoSAX (0.1-200 keV).
permits us to separate the main continuum components (``soft excess'', intrinsic power law,
reflection) with unprecedented precision.

Several homogeneous analyses of bright Seyfert 2s were published in the past years:
Turner \&
Pounds 1989 (EXOSAT observations, 0.1-10 keV), Nandra \& Pounds 1994 and Smith \& Done 1996 (GINGA
observations, 2-20 keV), Turner et al. 1997 (ASCA observations, 0.5-10 keV).
Many of the data analyzed in these works have been more carefully analyzed using specific models
for each source. The results are published in many papers where the X-ray emission of
a single or a couple of sources
is studied in detail.
However, the results obtained in these detailed studies are in most cases in agreement with the simpler
fits performed in the works quoted above where a large number ($>20$) of observations are analyzed.
The reason for this agreement is that even if peculiar features can be present in the X-ray spectrum of each
source, the statistics is dominated by a continuum  well fitted by a few components, namely
a thermal emission at low energies (from 0.1 to a few keV), an absorbed power law with photon index $\Gamma \sim 2$,
and a reflection component. Therefore, a homogeneous analysis of moderately large samples is
the best way to investigate the average X-ray properties of these sources, to test unified models and
to look for new correlations between the different spectral parameters.

The main results of the works quoted above support the unified models of AGNs, according to which
the narrow emission line (type 2) objects differ from broad emission line
objects (type 1) only for the presence of dusty absorbing gas along the line of sight.

The main X-ray spectral properties of Seyfert 2s can be modeled with the following 5 components:
\begin{enumerate}
\item A power-law with photon index $\Gamma \sim 2$. X-ray
observations of Seyfert 1s (Nandra et al. 1997) show that this power-law extends down to energies of a
fraction of keV. At high energies, an exponential cut-off is observed in the BeppoSAX observations
of a few objects (Guainazzi et al. 1999, Perola et al. 1999, 2000, Nicastro et al. 2000),
at energies from 60 to 300 keV. The physical origin of this component could
be a two phase accretion disk, where the soft photons emitted by a ``cold'' (kT $ < 50$ eV) thick disk
are Comptonized by a hot (kT$\sim 100$ keV) thin corona (Haardt \& Maraschi 1993).
\item A photoelectric cut-off of the intrinsic power-law is present at some energy, depending
of the column density, N$_{\rm H}$, of the absorbing gas. If 10$^{22}$ cm$^{-2} <$N$_{\rm H} <
10^{23}$ cm$^{-2}$, the cut-off energy is below $\sim 2$~keV, if  10$^{23}$ cm$^{-2} <$N$_{\rm H} <
10^{24}$ cm$^{-2}$ the cut-off energy is between 2 and 10 keV.
If N$_{\rm H} > 10  ^{24}$ cm$^{-2}$s$^{-1}$ no direct emission is observable below 10 keV. The intrinsic
component could only be revealed by observations in the 10-100 keV band.
If N$_{\rm H} > 10^{25}$ cm$^{-2}$s$^{-1}$  the source is completely thick at any energy, and only
reflected components (see below) are observable.
\item Thermal soft component: thermal emission, with kT ranging from 0.1 to a few keV is
a good description of the soft X-ray emission of most Seyfert 2s. This component in Seyfert 2s
cannot be due to emission from the accretion disk (since otherwise it would be obscured like the
power-law component), and is probably associated to the warm gas confining the broad emission
line clouds.
\item Reflected component: a cold reflected component is often required to obtain a good fit
of X-ray spectra of Seyfert galaxies above $\sim 5$ keV. The reflecting medium could be the accretion disk itself or,
alternatively, the inner edge of the absorbing gas (Ghisellini et al. 1994), or a wind (Elvis 2000).

A warm reflection component is also required in many cases. Assuming that this component is due to
scattering of the primary component with warm electrons, the spectral shape is the same as in the incident
continuum.

\item Iron line: an iron K$\alpha$ emission line, at energies E$\sim 6.4$  keV is observed in most Seyfert galaxies.
It is believed to be originated both in the accretion disk and in the absorbing/reflecting circumnuclear
gas (Matt et al. 1991, 1996) The observed equivalent width of the line is typically 100-300 eV in
Seyfert 1s and in Seyfert 2s with N$_{\rm H}$ lower than a few 10$^{23}$ cm$^{-2}$. In sources with heavier
absorption the equivalent width is higher (up to several keV) since the continuum at the line
energy is more absorbed than the line, that is -at least in part- emitted by the reflecting material along
free lines of sight. The line component emitted from far reflectors is narrow, and does not immediately
follow the continuum variations, while the disk component can be broad and is expected to vary with
almost zero delay with respect to the continuum.

\end{enumerate}

In this work we will use the components described above to fit the 0.1-200 keV observations
of a sample of 31 observations of 20 Seyfert 2s, performed by the BeppoSAX satellite in the period
1997-2000. 17 out of these
31 observations are currently unpublished.
In Sect. 2 we review the basic properties of the sample and we summarize the data reduction process.
In Sect. 3 we describe the models used to fit the data.
In Sect. 4 we discuss the results of data analysis. In Sect. 5 we summarize our conclusions.

\begin{table}
\centerline{ \begin{tabular}{ccc}
Name & z & Class.  \\
\hline
NGC 526a	 &  0.0295 & Sy 2  \\
NGC 1365     	 &  0.0055 & Sy 1.8  \\
IRAS 05189-2524  & 0.0426 & Sy 2  \\
NGC 2110            &  0.0076 & Sy 2  \\
NGC 2992             & 0.0077 & Sy 1.9  \\
MCG-5-23-16         & 0.0083 & Sy 2  \\
NGC 4258             & 0.0015 & Sy 1.9  \\
NGC 4388             & 0.0084 & Sy 2   \\
NGC 4507             & 0.0118 & Sy 2  \\
IRAS 13197-1627  & 0.0172 & Sy 1.8  \\
NGC 5252             & 0.0230 & Sy 1.9  \\
Centaurus A          & 0.0018 & Sy 2  \\
NGC 5506             & 0.0062 & Sy 1.9  \\
NGC 5674             & 0.0249 & Sy 1.9  \\
NGC 6300            & 0.0037  & Sy 2  \\
ESO 103-G35       & 0.0133 & Sy 2  \\
NGC 7172             & 0.0087 & Sy 2  \\
NGC 7314             & 0.0047 & Sy 1.9  \\
NGC 7582             & 0.0053 & Sy 2  \\
NGC 7679	 &  0.01714 & Sy 2 \\
\hline
\end{tabular}}
\caption{The sample of Compton thin, X-ray bright Seyfert 2 observed with BeppoSAX}
\end{table}
\section{Data: selection and reduction}

The sample consists of 31 observations of 20 sources (one source was observed 5 times, one
3 times and 5 sources were observed twice). The basic properties of the sample objects are listed
in Table 1.
\begin{table*}
\centerline{ \begin{tabular}{cccccc}
Name & Obs. date &  LECS exp. (s) & MECS exp. (s) &
PDS exp. (s) & Ref$^a$ \\
\hline
NGC 526a  	 & 31 DEC 98 & 38977 & 93175 & 44199 & 1 \\
NGC 1365    	 & 12 AUG 97 &  8862  & 27580 & 13023 & 2 \\
IRAS 05189-2524 & 03 OCT 99 & 17378 & 41973 & 19455 & 3 \\
NGC 2110    	 & 12 OCT 97 & 40665 & 83701 & 38673 & 4 \\
NGC 2992 \#1      & 01 DEC 97 & 26668 & 72007 & 33820 & 5 \\
NGC 2992 \#2      & 25 NOV 98 & 21914 & 59245 & 27084 & 5 \\
MCG-5-23-16   	 & 24 APR 98 & 35825 & 76990 & 32524 &  UNP.\\
NGC 4258	 & 19 DEC 98 & 32954 & 99428 & 46933 & 6 \\
NGC 4388 \#1 	 & 09 JAN 99 & 65242 & 67878 & 51189 & UNP. \\
NGC 4388 \#2  	 & 03 JAN 00 & 20822 & 28311 & 14607 & UNP. \\
NGC 4507 \#1 	 & 26 DEC 97 & 22229 & 55009 & 26967& UNP. \\
NGC 4507 \#2 	 & 02 JUL 98 &   8889  & 31408 & 16953 & UNP. \\
NGC 4507 \#3  	 & 13 JAN 99 & 11109 & 41325 & 20081 & UNP. \\
IRAS 13197-1627 & 22 JUL 98 & 14140 & 43780 & 18616 & UNP. \\
Centaurus A \#1 	 & 20 FEB 97 & 16912 & 33393 & 14873 & UNP. \\
Centaurus A \# 2  & 06 JAN 98  & 16681 & 54243 & 22987 & UNP. \\
Centaurus A \# 3   & 10 JUL 99 & 13676 & 38779 & 17801 & UNP. \\
Centaurus A \# 4   & 02 AUG 99 & 13670 & 38756 & 17801 & UNP. \\
Centaurus A \# 5   & 08 JAN 00  & 11775 & 34004 & 17941 & UNP. \\
NGC 5252    	 & 20 JAN 98 & 20897 & 61023 & 29155 & UNP. \\
NGC 5506 \#1  	 & 30 JAN 97 & 15495 & 39363 & 17008 & UNP. \\
NGC 5506 \#2  	 & 14 JAN 98 &   8868 & 39004 & 17500 & UNP. \\
NGC 5674	 & 12 FEB 00 & 21029 & 45039 & 20809 & UNP. \\
NGC 6300 	 & 28 AUG 99 & 39567& 86265 & 38555 & 7 \\
ESO 103-G35 \#1 & 03 OCT 96 &10088 & 50620 & 21029 & 8 \\
ESO 103-G35 \#2 &  14 OCT 97& 3617  & 14312 & 5915   & 8 \\
NGC 7172 \#1   	 & 14 OCT 96 & 15212 & 39167 & 17277 & 9 \\
NGC 7172 \#2   	 & 06 NOV 97 & 22405 & 49185 & 21147 & 9 \\
NGC 7314   	 & 08 JUN 99 & 35595 & 89799 & 42564 & UNP. \\
NGC 7582        	 & 09 NOV 98 & 26859 & 56435 & 26095 & 10 \\
NGC 7679         	 & 06 DEC 98 & 39799 & 90908 & 45298 & 11 \\
\hline
\end{tabular}}
\caption{$^a$ References: UNP:  previously unpublished; 1: Landi et al. 2001; 2: Risaliti et al. 2000; 3: Severgnini et al. 2000;
4: Malaguti et al. 1999; 5: Gilli et al. 2000; 6: Fiore et al. 2001; 7: Guainazzi 2002; 8: 
Wilkes et al. 2000, 9: Akylas et al. 2000;
10: Turner et al. 2000; 11: Della Ceca et al. 2001.}
\end{table*}

The only selection criteria are a 2-10 keV measured flux higher than
5$\times 10^{-12}$ erg cm$^{-2}$ s$^{-1}$, and a measured absorbing column density lower than
10$^{24}$ cm$^{-2}$ (i.e. ``Compton thin'' sources). As a consequence, the sample contains only
high signal-to-noise spectra, that can be studied in considerable detail.
The observation log is reported in Table 2. All the observations are available in the BeppoSAX
public archive provided by the ASI Science Data Center (SDC). Each source has
been observed with 3 narrow-field instruments on board of the Italian/Dutch BeppoSAX
satellite: LECS (0.1-10 keV, Parmar et al. 1995), MECS (1.65-10.5 keV, Boella
et al. 1995b) and PDS (20-200 keV, Frontera et al. 1995).

The LECS and MECS event files were obtained from the SDC public archive. The spectra were created
using the XSELECT v.2.0 code. A circular region of radius 4', centered on the source, was selected to extract
the spectrum.
The data were rebinned in order to have at least 20 counts per bin. This allows the use of Gaussian statistics
in the model fitting.
For the PDS data, we used the spectrum provided by the SDC public archive.
The background spectra for the LECS and MECS data were obtained from the long observations of blank
field provided by the SDC. Calibration matrices (redistribution matrix, RMF, for all the three instruments,
and ancillary (effective area) matrix, ARF, for LECS and MECS) were also provided by the SDC.

The spectra used in the scientific analysis consist of  LECS data from 0.1 to 3 keV (above 3 keV the
calibration of the LECS is uncertain, and we do not lose much in terms of statistics, since the area of
the MECS is significantly higher than that of the LECS above 2 keV); MECS data from 1.65 to 10.5 keV,
and PDS data from 20 to 200 keV (in several cases, where the statistics at high energy was too low,
we cut the spectrum at 100 keV).

\section{Model fitting}

The data analysis was performed using the XSPEC v.10.0 code. All the errors quoted in this works
are at a confidence level of 90\% for one interesting parameter.

{In all models we fixed the normalization factor between PDS and MECS to 0.8.
The uncertainty in this factor is at most 10\%, according to the SAX SDC. This
is not a dominant source of error, since te typical final errors in the
parameters strongly affected by high-energy data are typically much higher.
This is due to both the limited number of counts and the degeneracy between
some of the spectral components, as will be further discussed in the following
Subsections. The normalization between LECS and MECS were left free. The
typical best
fit values are in the interval 0.6-0.8.

The Galactic absorption is taken into account in all models, by multiplying
each component by an extra
absorption
factor. }
 
First, we fitted the data with a simple model, consisting of an absorbed power law plus an unabsorbed
power law. This model is a good zero-order approximation of the continuum in Seyfert 2s. Our main aim
is to highlight the other spectral features from the analysis of the residuals.
In Table 3 we show the results of the spectral analysis. In Fig. \ref{fig:s1} we plot for each spectrum the 
folded
data+model and the contribution to $\chi^2$ from each spectral bin. This is an effective way to look
graphically for statistically significant unfitted features.
\begin{table*}
\centerline{ \begin{tabular}{ccccccc}
Name& Flux$^a$&  $\Gamma_1^b$
& N$_{\rm H}^c$ & $\Gamma_2^d$
& R$^e$ & $\chi^2$/d.o.f.\\
\hline
NGC 526a        &1.72 &1.55$^{+0.07}_{-0.05}$ &1.6$_{-0.2}^{+0.2}$ &1.3$^{+0.6}_{-0.3}$   &
38  & 278/242 \\
NGC 1365        &0.68 &1.58$_{-0.16}^{+0.21}$ &44$_{-8}^{+8}$ &1.8$_{-0.3}^{+0.3}$   &
11 &129/77 \\
IRAS 05189-2524 &0.35 &2.41$_{-0.30}^{+0.26}$ &8.6$_{-1.7}^{+2.0}$ &2.1$_{-0.6}^{+0.5}$   &
218 &88/72   \\
NGC 2110        &3.00 &1.65$_{-0.04}^{+0.05}$ &3.9$_{-0.2}^{+0.3}$ &1.48$_{-0.25}^{+0.45}$ &
24 &358/282 \\
NGC 2992 \#1    &0.61 &1.46$_{-0.08}^{+0.07}$ &0.66$_{-0.11}^{+0.13}$ & --                &
--  &253/174 \\
NGC 2992 \#2    &7.37 &1.69$_{-0.02}^{+0.02}$ &0.85$_{-0.04}^{+0.05}$ &--               &
--  & 402/349 \\
MCG-5-23-16     &9.28 &1.72$_{-0.02}^{+0.02}$ &1.63$_{-0.05}^{+0.05}$ & --                &
--  & 536/376 \\
NGC 4258        &0.79 &2.33$_{-0.15}^{+0.17}$ &13.3$_{-1.3}^{+1.5}$ &1.85$_{-0.11}^{+0.10}$ &
23 & 266/197 \\
NGC 4388 \#1    &2.53 &1.60$_{-0.03}^{+0.03}$ &41.4$_{-1.7}^{+1.9}$ &1.73$_{-0.11}^{+0.10}$ &
42 & 382/239 \\
NGC 4388 \#2    &0.94 &1.58$_{-0.12}^{+0.14}$ &54$_{-7}^{+9}$ &1.75$_{-0.25}^{+0.23}$ &
18 & 166/105 \\
NGC 4507 \#1    &1.84 &1.68$_{-0.05}^{+0.05}$ &64$_{-4}^{+4}$      &1.2$_{-0.2}^{+0.2}$   &
174 & 334/186 \\
NGC 4507 \#2    &1.65 &1.66$_{-0.08}^{+0.07}$ &59$_{-5}^{+5}$      &0.93$_{-0.25}^{+0.35}$&
260 & 230/138 \\
NGC 4507 \#3    &0.87 &1.50$_{-0.10}^{+0.15}$ &64$_{-10}^{+9}$     &1.1$_{-0.3}^{+0.3}$   &
63  & 258/127 \\
IRAS 13197-1627 &0.47 &1.70$_{-0.28}^{+0.33}$ &33$_{-6}^{+7}$ &3.5$_{-0.9}^{+1.2}$     &
6.5 &144/87 \\
Centaurus A  \#1   &19.6 &1.79$_{-0.02}^{+0.01}$ &10.1$_{-0.2}^{+0.3}$ &1.86$_{-0.24}^{+0.23}$ &
56 & 417/259 \\
Centaurus A  \#2   &25.4 &1.77$_{-0.01}^{+0.01}$ &9.40$_{-0.15}^{+0.16}$ &2.05$_{-0.25}^{+0.23}$ &
67 & 396/273 \\
Centaurus A  \#3   &23.6 &1.78$_{-0.01}^{+0.01}$ &9.83$_{-0.20}^{+0.21}$ &1.89$_{-0.27}^{+0.25}$ &
65 & 403/255 \\
Centaurus A  \#4   &23.6 &1.78$_{-0.01}^{+0.01}$ &9.86$_{-0.19}^{+0.19}$ &2.10$_{-0.23}^{+0.20}$ &
54 & 407/266 \\
Centaurus A  \#5   &23.1 &1.77$_{-0.01}^{+0.02}$ &9.93$_{-0.23}^{+0.24}$ &1.53$_{-0.20}^{+0.23}$ &
70 & 374/247 \\
NGC 5252        &0.26 &1.52$_{-0.30}^{+0.30}$ &6.2$_{-1.3}^{+1.8}$ &3.7$_{-2.4}^{+2.2}$ &
29 &80/70 \\
NGC 5506 \#1    &7.47 &1.82$_{-0.04}^{+0.05}$ &3.46$_{-0.15}^{+0.17}$ &1.2$_{-0.2}^{+0.3}$ &
96 &408/256  \\
NGC 5506 \#2    &7.21 &1.75$_{-0.03}^{+0.03}$ &3.41$_{-0.18}^{+0.16}$ &1.7$_{-0.5}^{+4.0}$ &
76 &340/228  \\
NGC 5764 	& 0.54 & 1.70$_{-0.19}^{+0.21}$ & 6.8$_{-1.3}^{+1.7}$ &  2.9$_{-1.0}^{+0.9}$  &
13 & 91/107  \\
NGC 6300 	& 1.29 & 1.71$_{-0.07}^{+0.07}$ & 21.9$_{-1.2}^{+1.2}$ & 2.8$_{-0.3}^{+0.3}$ &
78 & 261/183 \\
ESO 103-G35 \#1 &2.63 &1.94$_{-0.05}^{+0.06}$ &20.5$_{-0.8}^{+0.8}$&--                    &
--  & 255/183 \\
ESO 103-G35 \#2 &1.53 &1.73$_{-0.14}^{+0.18}$ &20$_{-2}^{+4}$      &--                    &
--  & 56/76   \\
NGC 7172 \#1    &1.12 &1.68$_{-0.11}^{+0.11}$ &10.5$_{-1}^{+1}$    &2.6$_{-0.6}^{+0.7}$   &
123 & 152/165 \\
NGC 7172 \#2    &0.60 &1.59$_{-0.15}^{+0.17}$ &9.9$_{-1.2}^{+1.6}$ &2.3$_{-0.7}^{+0.5}$   &
46  & 153/161 \\
NGC 7314        &2.39 &1.94$_{-0.08}^{+0.07}$ &1.08$_{-0.13}^{+0.12}$ &1.03$_{-0.09}^{+0.07}$ &
25 & 339/288 \\
NGC 7582        &2.0  &1.55$_{-0.05}^{+0.05}$ &12.8$_{-0.8}^{+0.8}$ &2.1$_{-0.3}^{+0.3}$    &
36  & 265/187 \\
NGC 7679        &0.59 &1.95$_{-0.20}^{+0.23}$ &0.10$_{-0.04}^{+0.07}$ &	1.0$_{-0.8}^{+0.3}$&
14 &207/222 \\
\hline
\end{tabular}}
 \caption{\footnotesize{Model A: Absorbed power-law plus a second unabsorbed power-law. Since this second
component fits mainly the soft excess, the ratio between the normalizations of
the two components gives an idea of the relative strength of the soft excess
with respect to the primary component.
$^a$ Flux 2-10 keV in units of 10$^{11}$ erg s$^{-1}$ cm$^{-2}$;
$^b$ $\Gamma_1$: Photon index of the absorbed power-law;
$^c$ N$_{\rm H}$: X-ray absorbing column density, in units of 10$^{22}$ cm$^{-2}$;
$^d$ $\Gamma_2$: Photon index of the transmitted power-law;
$^e$ R: Ratio between the normalizations of the absorbed and transmitted components.
}}
\end{table*}

From the analysis of Fig. \ref{fig:s1} we can identify the following unfitted features:
\begin{itemize}
\item At low energies (E$<$2 keV) several sources present a more complex spectral shape than a simple
power-law. This is apparent, for example, for the objects NGC~526a, NGC~4258, NGC~4388, NGC~4507, Cen~A.
\item An excess around 6.5 keV is present in most observations.
\item The high energy spectrum (E$>$ 20 keV) is in many cases convex, and cannot be
reproduced by a single power-law.
\end{itemize}
We also note that in all the sources but one (NGC 7679) the photoelectric cut-off is highly significant,
and permits a precise determination of the absorbing column density, N$_{\rm
H}$.
\begin{table*}
\centerline{ \begin{tabular}{ccccccccc}
Name &  $\Gamma^a$
& N$_{\rm H}^b$ & E$_{\rm C}^c$ & kT$^d$
& E$_{\rm Fe}^e$ & W$_{\rm Fe}^f$ &
EW$_{\rm Fe}^g$ & $\chi^2$/d.o.f. \\
\hline
NGC 526a        &1.54$^{+0.05}_{-0.04}$ &1.6$_{-0.2}^{+0.2}$     &$>$ 176 & $>$1.1 &
6.65$^{+0.17}_{-0.21}$ & 0.33$_{-0.33}^{+0.21}$ & 182$_{-96}^{+76}$ & 251/238 \\
NGC 1365        & 0.75$_{-0.45}^{+0.23}$ &26$_{-8}^{+9}$ &  44$_{-20}^{+42}$ &2.7$_{-1.1}^{+3.0}$ &
6.22$_{-0.10 }^{+0.10}$ & -- &  642$_{-230}^{+248}$ &106/74 \\
IRAS 05189-2524 &2.33$_{-0.30}^{+0.26}$ &7.6$_{-1.0}^{+1.5}$ & $>15 $ & 0.88$_{-0.23}^{+0.44}$ &
6.47$_{-0.20}^{+0.23}$ & $<0.45$ & 239$_{-153}^{+187}$ &  86/68   \\
NGC 2110        &1.62$_{-0.07}^{+0.07}$ & 3.7$_{-0.2}^{+0.3}$ & $>166$ &  $>2.6$ &
6.34$_{-0.08}^{+0.07}$ & $<0.32$& 235$_{-60}^{+63}$ & 275/278 \\
NGC 2992 \#1    & 1.63$_{-0.14}^{+0.08}$& 1.1$_{-0.4}^{+0.5}$& $>112$  & 1.1$_{-0.2}^{+1.2}$ &
6.61$_{-0.07}^{+0.06}$ & 0.16$_{-0.08}^{+0.11}$ & 684$_{-132}^{+158}$ & 172/170 \\
NGC 2992 \#2    & 1.71$_{-0.04}^{+0.02}$& 0.85$_{-0.06}^{+0.05}$& $>215$ & -- &
6.55$_{-0.11}^{+0.11}$& 0$^m$& 103$_{-28}^{+28}$  & 359/346 \\
MCG-5-23-16     &  1.69$_{-0.03}^{+0.03}$& 1.54$_{-0.06}^{+0.07}$ & 261$_{-88}^{+252}$ & -- &
6.44$_{-0.06}^{+0.06}$&  -- & 108$_{-22}^{+22}$  &  440/373 \\
NGC 4258    & 2.25$_{-0.05}^{+0.03}$ & 12.1$_{-0.6}^{+0.7}$ & $>145$& 0.67$_{-0.07}^{+0.74}$ &
6.56$_{-0.19}^{+0.21}$ & $<0.35$ & 80$_{-62}^{+68}$ & 162/192 \\
NGC 4388 \#1 & 1.23$_{-0.10}^{+0.08}$& 31.4$_{-2.5}^{+2.0}$  & 108$_{-28}^{+42}$ &  2.4$_{-0.6}^{+0.6}$ &
6.42$_{-0.07}^{+0.07}$ & 0.28$_{-0.14}^{+0.09}$ & 606$_{-131}^{+154}$ & 272/236  \\
NGC 4388 \#2 & 1.25$_{-0.11}^{+0.18}$ &38$_{-3}^{+4}$ &  264 $_{-109}^{+400}$ & 3.4$_{-0.9}^{+2.2}$ &
6.36$_{-0.05}^{+0.05}$& $<0.12$& 564$_{-120}^{+120}$ & 111/101 \\
NGC 4507 \#1    & 0.69$_{-0.16}^{+0.17}$ & 38.5$_{-2.5}^{+4.8}$ & 36$_{-8}^{+9}$ &  3.7$_{-1.7}^{+5.1}$ &
6.35$_{-0.10}^{+0.09}$& 0.43$_{-0.12}^{+0.12}$ & 1150$_{-220}^{+253}$ & 234/182 \\
NGC 4507 \#2    & 0.56$_{-0.31}^{+0.24}$ & 34.6$_{-4.4}^{+3.3}$ & 35$_{-3}^{+13}$ & $>2.4$  &
6.44$_{-0.09}^{+0.08}$ & 0.44$_{-0.11}^{+0.13}$ &  940$_{-243}^{+278}$ & 150/134 \\
NGC 4507 \#3    & 0.1$_{-0.4}^{+0.2}$ & 20.4$_{-8.9}^{+7.1}$& 27.1$_{-8.6}^{+8.8}$ & 2.3$_{-1.0}^{+4.5}$&
6.44$_{-0.06}^{+0.06}$ &0.23$_{-0.08}^{+0.09}$ & 1000$_{-200}^{+270}$ & 164/124 \\
IRAS 13197-1627 &1.31$_{-0.23}^{+0.08}$ &24.4$_{-3.3}^{+3.5}$ & $>106$ & 1.0$_{-0.2}^{+0.3}$&
6.40$_{-0.08}^{+0.09}$ & 0$^m$ & 463$_{-140}^{+128}$ & 116/77 \\
Centaurus A \#1 &1.72$_{-0.04}^{+0.03}$ &  9.70$_{-0.26}^{+0.25}$ & $>231$ & 6.7$_{-2.2}^{+4.7}$ &
6.40$_{-0.06}^{+0.06}$ &  0$^m$ & 98$_{-17}^{+16}$  & 297/256 \\
Centaurus A \#2 &1.70$_{-0.03}^{+0.02}$ & 9.07$_{-0.09}^{+0.17}$ & 351$_{-94}^{+170}$ & 4.7$_{-1.4}^{+2.4}$ &
6.38$_{-0.08}^{+0.07}$ &  0$^m$ & 50$_{-14}^{+13}$  & 342/270 \\
Centaurus A \#3 & 1.73$_{-0.03}^{+0.02}$ &9.46$_{-0.22}^{+0.21}$ & $>336$ & 5.9$_{-2.2}^{+3.6}$ &
6.52$_{-0.05}^{+0.04}$ &  0$^m$ & 109$_{-18}^{+17}$  & 271/252 \\
Centaurus A \#4 & 1.73$_{-0.03}^{+0.03}$ & 9.54$_{-0.23}^{+0.22}$ & $>345$ &4.0$_{-1.2}^{+1.0}$  &
6.52$_{-0.04}^{+0.05}$ & 0$^m$ &108$_{-18}^{+18}$ & 291/263 \\
Centaurus A \#5 & 1.83$_{-0.01}^{+0.02}$ & 10.1$_{-0.2}^{+0.2}$ & $>588$ & 15$_{-6}^{+12}$  &
6.46$_{-0.05}^{+0.05}$ & 0$^m$ & 118$_{-20}^{+19}$ & 307/244 \\
NGC 5252          &1.68$_{-0.34}^{+0.19}$ &6.1$_{-1.4}^{+1.4}$ &  $>58$& $>8.6 $ &
6.48$_{-0.16}^{+0.15}$ & $<0.24$ &267$_{-161}^{+158}$ & 76/66 \\
NGC 5506 \#1    & 1.82$_{-0.03}^{+0.02}$& 3.28$_{-0.12}^{+0.13}$ & $>415$ & 0.50$_{-0.21}^{+0.60}$ &
6.59$_{-0.08}^{+0.08}$& 0.26$_{-0.09}^{+0.12}$ & 199$_{-38}^{+50}$ & 321/253 \\
NGC 5506 \#2    &1.68$_{-0.06}^{+0.04}$ &3.15$_{-0.21}^{+0.14}$ & 214$_{-75}^{+332}$ & 2.3$_{-2.2}^{+4.4}$&
6.48$_{-0.08}^{+0.08}$ & $<0.22$ &130$_{-34}^{+46}$ & 282/224  \\
NGC 5764 &1.69$_{-0.14}^{+0.21}$ & 6.3$_{-0.7}^{+1.3}$ & $>340$ & 1.1$_{-0.3}^{+2.3}$ &
6.36$_{-0.20}^{+0.15}$ &  $<0.47$&  221$_{-108}^{+180}$& 82/103 \\
NGC 6300 &1.62$_{-0.12}^{+0.07}$  & 20.7$_{-1.1}^{+1.2}$  & $>184$ & 1.0$_{-0.1}^{+1.7}$  &
6.4$^m$ &0.44$_{-0.25}^{+0.26}$ & 267$_{-161}^{+158}$ & 236/180 \\
ESO 103-G35 \#1 &1.69$_{-0.16}^{+0.13}$ &18.2$_{-1.5}^{+0.6}$    &84$_{-19}^{+104}$ &-- &
6.44$_{-0.12}^{+0.08}$ & 0.20$_{-0.10}^{+0.20}$ & 234$_{-34}^{+124}$ & 185/181 \\
ESO 103-G35 \#2 &1.62$_{-0.32}^{+0.19}$ & 18.3$_{-2.4}^{+2.6}$ & $>67$ & $>0.22$ &
6.49$_{-0.19}^{+0.12}$ & $<0.35$ &244$_{-125}^{+119}$  & 44/72   \\
NGC 7172 \#1    & 1.64$_{-0.14}^{+0.12}$ & 9.7$_{-0.8}^{+0.9}$& $>109$ &  $<1.1$ &
6.46$_{-0.13}^{+0.15}$ & $<0.6$ &116$_{-65}^{+127}$  & 148/162 \\
NGC 7172 \#2    &1.72$_{-0.11}^{+0.08}$ &10.8$_{-1.1}^{+0.8}$     &  $>103$ &  $>2.4$&
6.62$_{-0.16}^{+0.09}$ & $<0.33$ & 157$_{-81}^{+121}$ & 149/157 \\
NGC 7314        &1.79$_{-0.03}^{+0.05}$ &0.86$_{-0.06}^{+0.08}$  &$>504$  &  0.18$_{-0.03}^{+0.05}$&
6.49$_{-0.13}^{+0.13}$ & $<0.42 $&  169$_{-75}^{+56}$&
330/284 \\
NGC 7582        &1.31$_{-0.03}^{+0.03}$ &10.5$_{-0.4}^{+0.4}$  & 109$_{-20}^{+30}$ & 1.04$_{-0.08}^{+0.08}$ &
6.28$_{-0.21}^{+0.20}$ & 0$^m$ &  112$_{-51}^{+45}$ &  243/184 \\
NGC 7679        &1.70$_{-0.10}^{+0.07}$ & $<0.05$ & $>90$ &1.1$_{-0.2}^{+0.9}$ &
--  & -- & -- & 208/221 \\
\hline
\end{tabular}}
\caption{\footnotesize{Model B: Absorbed power-law, Raymond-Smith component, iron line.
$^a$ $\Gamma$: Photon index of the absorbed power-law;
$b$ N$_{\rm H}$: X-ray absorbing column density, in units of 10$^{22}$ cm$^{-2}$;
$^c$ E$_{\rm C}$: Exponential cut-off energy (keV);
$^d$ kT: Temperature of the thermal component in units of kT (keV);
$^e$ E$_{\rm Fe}$: Fe K$\alpha$ line energy (keV);
$^f$ W$_{\rm Fe}$: Line width (keV);
$^g$ EW$_{\rm Fe}$: Line equivalent width (eV).
$^m$ Fixed parameter. }}
\end{table*}

At high energies the observed curvature can have two different
origins: an high energy exponential cut-off, or a reflection component.
In this case a degeneracy between these two components is possible, so we fitted our data with
two new models: in the first one (model B) we replaced the unabsorbed power-law with a Raymond-Smith
thermal model, and we added a Gaussian and an exponential cut-off to the absorbed power-law.
In the second model (model C) we also added a warm reflection component, using a power-law with the photon
index equal to that of the primary continuum, and a cold reflection component, using the PEXRAV model in
XSPEC. This model assumes an infinite plane geometry for the reflector. This may not be a good
approximation of the real reflector. However, the overall spectral shape of the reflection component should
be only weakly dependent on the details of the absorber. The dependence
of the spectral shape on the reflection angle in the PEXRAV model is low, therefore,
given the uncertainty in the real absorber geometry, we chose
to fix the inclination angle in the PEXRAV model to the value of $\theta=$30 deg
(the value has been chosen in order
to have an average ratio between the observed section of the absorber and it real area of $\cos \theta=0.5$).
The results of the fits with model C are summarized in Table 5.
\begin{table*}
\centerline{ \begin{tabular}{ccccccccccc}
Name &  $\Gamma^a$
& N$_{\rm H}^b$ & E$_{\rm C}^c$ & kT$^d$
& E$_{\rm Fe}^e$ & W$_{\rm Fe}^f$ &
EW$_{\rm Fe}^g$ & R$_{\rm C}^h$&R$_{\rm W}^h$ & $\chi^2$/d.o.f. \\
\hline
NGC 526a        &1.54$_{-0.04}^{+0.05}$ &1.6$_{-0.2}^{+0.2}$  &$>$ 176 		  & $>$1.1 &
6.65$_{-0.21}^{+0.17}$ & 0.33$_{-0.33}^{+0.21}$ & 182$_{-96}^{+76}$ &$<0.5$ &  $<0.1$ & 251/237 \\
NGC 1365        &1.74$_{-0.39}^{+0.21}$ &39$_{-12}^{+20}$ & $>56$  & 1.0$_{-0.4}^{+0.6}$ &
6.24$_{-0.12}^{+0.11}$ &  0$^m$ & 290$_{-139}^{+129}$  & 3.8$_{-2}^{+2}$  & $<0.1$ & 91/72 \\
IRAS 05189-2524 &2.59$_{-0.43}^{+0.35}$ &8.6$_{-0.7}^{+1.5}$ & $>25$ & --  &
6.46$_{-0.25}^{+0.27}$ & $<0.45$ & 256$_{-192}^{+210}$ & $<4$ & 0.02$_{-0.01}^{+0.02}$ & 78/68 \\
NGC 2110        &1.66$_{-0.09}^{+0.08}$ &3.8$_{-0.3}^{+0.3}$ & $> 143$ &$>2.9$ &
6.38$_{-0.08}^{+0.08}$ & $<0.33$ & 242$_{-72}^{+61}$  & $<0.4$ &$<0.03$& 274/277 \\
NGC 2992 \#1    &1.77$_{-0.16}^{+0.18}$ &1.4$_{-0.4}^{+0.4}$  & $>53$  & 1.16$_{-0.29}^{+1.06}$ &
6.63$_{-0.08}^{+0.06}$ & $<0.28$ & 765$_{-86}^{+160}$  &0.8$_{-0.6}^{+4}$   &$<0.06$ & 170/169 \\
NGC 2992 \#2    &1.79$_{-0.03}^{+0.08}$ &0.93$_{-0.03}^{+0.08}$ &  $>141$ & -- &
6.52$_{-0.33}^{+0.14}$ & 0.36$_{-0.12}^{+0.20}$ & 154$_{-33}^{+47}$ & 0.3$_{-0.1}^{+0.4}$ &$<0.01$& 350/344 \\
MCG-5-23-16     &1.77$_{-0.04}^{+0.04}$ &1.59$_{-0.04}^{+0.04}$& 157$_{-34}^{+81}$ & -- &
6.45$_{-0.07}^{+0.08}$&$<0.30$ &126$_{-24}^{+31}$ & 0.54$_{-0.17}^{+0.26}$ &$<0.01$& 404/373 \\
NGC 4258        &2.25$_{-0.05}^{+0.04}$ &12.1$_{-0.6}^{+0.7}$ & $>145$ &0.67$_{-0.07}^{+0.07}$  &
6.56$_{-0.28}^{+0.30}$ & $<0.35$ & 80$_{-40}^{+70}$ & $<0.6$ & $<0.02$& 162/191 \\
NGC 4388 \#1    &1.58$_{-0.22}^{+0.08}$ &38$_{-4}^{+2}$ & $>145$ &1.40$_{-0.06}^{+0.25}$  &
6.46$_{-0.10}^{+0.07}$ & $<0.23$ & 233$_{-35}^{+115}$ & 0.3$_{-0.1}^{+0.2}$ &0.016$_{-0.003}^{+0.004}$ & 236/233 \\
NGC 4388 \#2    &1.47$_{-0.41}^{+0.04}$ &48$_{-8}^{+18}$ & $>109$ &1.07$_{-0.22}^{+0.28}$  &
6.38$_{-0.06}^{+0.05}$ & $<0.12$ & 525$_{-112}^{+115}$ & $<1.4$ &0.06$_{-0.02}^{+0.02}$& 100/99 \\
NGC 4507 \#1    &1.51$_{-0.05}^{+0.06}$ &59$_{-12}^{+8}$       & 126$_{-48}^{+153}$ & 0.87$_{-0.14}^{+0.20}$ &
6.40$_{-0.19}^{+0.10}$ & 0.33$_{-0.16}^{+0.14}$ & 297$_{-100}^{+109}$ &0.7$_{-0.2}^{+0.2}$   &$<0.01$& 206/180 \\
NGC 4507 \#2    &1.56$_{-0.15}^{+0.05}$ &54$_{-7}^{+9}$       &$>100$             &0.71$_{-0.40}^{+0.27}$ &
6.47$_{-0.12}^{+0.11}$ & 0.37$_{-0.16}^{+0.18}$ &520$_{-163}^{+557}$ & 0.9$_{-0.3}^{+0.3}$ &$<0.01$& 132/133 \\
NGC 4507 \#3    &1.70$_{-0.13}^{+0.04}$ &71$_{-16}^{+20}$     &$>216$ &0.78$_{-0.26}^{+0.22}$ &
6.43$_{-0.07}^{+0.06}$ & $<0.25$ & 433$_{-29}^{+242}$ & 2.0$_{-0.5}^{+0.5}$ &$<0.02$& 132/122 \\
IRAS 13197-1627 &1.77$_{-0.07}^{+0.07}$ &33$_{-4}^{+5}$          &$>164$ &0.90$_{-0.10}^{+0.13}$  &
6.43$_{-0.08}^{+0.10}$ & $<0.27$ & 585$_{-190}^{+275}$ & 2.1$_{-0.6}^{+0.5}$  &$<0.01$& 98/77 \\
Centaurus A \#1  &1.81$_{-0.03}^{+0.05}$ &10.0$_{-0.2}^{+0.3}$ & $>429 $ &--  &
6.48$_{-0.07}^{+0.08}$ & 0.33$_{-0.11}^{+0.12}$ & 173$_{-30}^{+39}$ & 0.18$_{-0.07}^{+0.19}$& 0.015$_{-0.002}^{+0.002}$& 259/254 \\
Centaurus A  \#2 &1.78$_{-0.02}^{+0.03}$ &9.30$_{-0.17}^{+0.13}$ & $>490$ &0.56$_{-0.28}^{+0.15}$  &
6.41$_{-0.09}^{+0.09}$ & $<0.5$ & 75$_{-27}^{+36}$ & 0.15$_{-0.05}^{+0.07}$ & 0.011$_{-0.002}^{+0.002}$& 290/267 \\
Centaurus A  \#3 &1.73$_{-0.01}^{+0.01}$ &9.43$_{-0.08}^{+0.23}$ & $>339$ &--  &
6.57$_{-0.04}^{+0.05}$ & $<0.14$ & 138$_{-21}^{+14}$ & $<0.07$ & 0.015$_{-0.002}^{+0.002}$& 259/250 \\
Centaurus A  \#4 &1.73$_{-0.01}^{+0.02}$ &9.36$_{-0.18}^{+0.17}$ & 530$_{-186}^{+205}$ &0.54$_{-0.26}^{+0.15}$  &
6.57$_{-0.08}^{+0.02}$ & $<0.19$ & 141$_{-21}^{+17}$ & $<0.08$ &0.014$_{-0.002}^{+0.002}$& 245/260 \\
Centaurus A \#5 &1.75$_{-0.01}^{+0.01}$ &9.43$_{-0.11}^{+0.14}$ & $>605 $ &0.47$_{-0.24}^{+6.22}$  &
6.45$_{-0.06}^{+0.04}$ & $<0.11$ & 117$_{-16}^{+20}$ & 0.09$_{-0.02}^{+0.02}$ &0.014$_{-0.002}^{+0.002}$& 255/241 \\
NGC 5252        &1.83$_{-0.12}^{+0.13}$ &6.8$_{-0.7}^{+1.6}$ & $>49$ & 0.54$_{-0.34}^{+0.47}$ &
6.47$_{-0.17}^{+0.15}$ & $<0.25$ & 278$_{-171}^{+215}$ & 1.6$_{-1.2}^{+1.2}$ &$<0.01$&71/65 \\
NGC 5506 \#1    &2.03$_{-0.03}^{+0.02}$ &3.68$_{-0.09}^{+0.17}$ & $>398$ & $>0.27$  &
6.70$_{-0.14}^{+0.09}$ & 0.36$_{-0.12}^{+0.15}$ &229$_{-58}^{+82}$  &2.3$_{-1}^{+1}$  &$<0.01$ & 269/251 \\
NGC 5506 \#2    &2.02$_{-0.02}^{+0.02}$ &3.85$_{-0.19}^{+0.10}$  & $>314$  & 0.84$_{-0.5}^{+1.0}$ &
6.52$_{-0.11}^{+0.15}$& $<0.25$ & 164$_{-32}^{+103}$ &  1.2$_{-0.2}^{+0.2}$ &$<0.03$& 224/220 \\
NGC 5674 & 1.82$_{-0.19}^{+0.09}$& 6.5$_{-0.5}^{+0.8}$ & $>112$  & 1.1$_{-0.7}^{+1.7}$ &
6.36$_{-0.37}^{+0.17}$ & $<0.5 $ & 236$_{-124}^{+219}$ & 1$_{-0.7}^{+2}$ &$<0.06$& 79/102 \\
NGC 6300 & 1.87$_{-0.04}^{+0.04}$ & 22.8$_{-0.9}^{+0.9}$ & $>384$ & 0.82$_{-0.09}^{+0.07}$ &
6.27$_{-0.12}^{+0.12}$ & 0$^m$ & 150$_{-44}^{+45}$ & 1.1$_{-0.2}^{+0.2}$&$<0.01$& 197/179 \\
ESO 103-G35 \#1 &1.85$_{-0.18}^{+0.09}$ &19.7$_{-1.7}^{+0.6}$ &87$_{-33}^{+134}$  &-- &
6.37$_{-0.09}^{+0.13}$ & 0.22$_{-0.05}^{+0.16}$ & 215$_{-62}^{+79}$ &0.4$_{-0.2}^{+0.2}$  & $<0.01$&  172/180 \\
ESO 103-G35 \#2 &1.85$_{-0.33}^{+0.13}$ &20.5$_{-3.3}^{+3.1}$ &$>84$  		  & -- &
6.49$_{-0.15}^{+0.12}$ & $<0.33$                & 224$_{-116}^{+126}$& 0.7$_{-0.6}^{+0.8}$ &$<0.01$& 42/73   \\
NGC 7172 \#1    &1.88$_{-0.10}^{+0.06}$ &11.1$_{-1.2}^{+1.2}$ & $>128$ &$<0.44$ &
6.4$^m$                & 0$^m$                 & 90$_{-70}^{+70}$ & 1.1$_{-0.9}^{+2}$ &$<0.01$& 142/162 \\
NGC 7172 \#2    &1.97$_{-0.08}^{+0.09}$ &11.0$_{-0.8}^{+0.9}$   &$>105$ & 0.19$_{-0.08}^{+0.04}$ &
6.55$_{-0.13}^{+0.17}$ & $<0.31$ & 199$_{-111}^{+144}$ & 2.4$_{-0.9}^{+0.8}$ &$<0.01$& 140/156 \\
NGC 7314        &2.07$_{-0.05}^{+0.02}$ &1.22$_{-0.14}^{+0.09}$ & $>462$  & 3.7$_{-1.3}^{+4.0}$ &
6.37$_{-0.17}^{+0.18}$ & $<0.45$ & 102$_{-51}^{+88}$ & 2.4$_{-1.0}^{+0.4}$ &$<0.04$& 276/283 \\
NGC 7582        &1.93$_{-0.03}^{+0.03}$ &14.3$_{-0.5}^{+0.7}$ &$>364$ & 0.78$_{-0.12}^{+0.09}$ &
6.4$^m$ & 0$^m$ & 183$_{-125}^{+269}$ & 2.2$_{-0.2}^{+0.6}$& $<0.01$& 175/184 \\
NGC 7679        &2.08$_{-0.21}^{+0.24}$ &0.6$_{-0.4}^{+0.05}$  & $>38$ &-- &
-- & -- & -- & 2.7$_{-1.8}^{+3.6}$ & --& 198/220 \\
\end{tabular}}
\caption{\footnotesize{Model C: Absorbed power-law, Raymond Smith component, cold reflection, warm reflection,
iron line.
$^a$ $\Gamma$: Photon index of the absorbed power-law;
$^b$ N$_{\rm H}$: X-ray absorbing column density, in units of 10$^{21}$ cm$^{-2}$;
$^c$ E$_{\rm C}$: Exponential cut-off energy (keV);
$^d$ kT: Temperature of the thermal component in units of kT (keV);
$^e$ E$_{\rm Fe}$: Fe K$\alpha$ line energy (keV);
$^f$ W$_{\rm Fe}$: Line width (keV);
$^g$ EW$_{\rm Fe}$: Line equivalent width (eV);
$^h$ R$_{\rm C}$, R$_{\rm W}$: Ratio between the normalizations of the reflected and direct components.
$^m$ Fixed parameter.}}
\end{table*}

\section{Results and Discussion}

Our models A,B,C provide a good fit to almost all the sources of the sample: we obtained a reduced $\chi^2$
lower than 1.10 in 29 cases out of 31. The best model for each observation, and the reduced $\chi^2$
are listed in Table 6. We briefly comment about the ``bad'' fits ($\chi^2 >$ 1.1) in the
Appendix. Here we  discuss
the main results of our analysis:

\begin{table}
\centerline{\begin{tabular}{ccc}
Name & Mod. & $\chi^2_{\rm r}$ \\
\hline
NGC 526a  	 & B & 1.05 \\
NGC 1365    	 & C & 1.26 \\
IRAS 05189-2524 & C &  1.15 \\
NGC 2110    	 & B  & 0.99 \\
NGC 2992 \#1      &  B & 1.01 \\
NGC 2992 \#2      &  C & 1.02 \\
MCG-5-23-16   	 & C & 1.08 \\
NGC 4258	 & B & 0.84 \\
NGC 4388 \#1 	 & C & 1.01 \\
NGC 4388 \#2  	 & C & 1.01  \\
NGC 4507 \#1 	 & C & 1.08  \\
NGC 4507 \#2 	 & C & 0.99  \\
NGC 4507 \#3  	 & C & 1.08  \\
IRAS 13197-1627 & C & 1.27  \\
Centaurus A \#1 	 & C & 1.02  \\
Centaurus A \# 2  &  C & 1.09 \\
Centaurus A \# 3   & C & 1.04  \\
Centaurus A \# 4   & C & 0.94  \\
Centaurus A \# 5   & C & 1.10  \\
NGC 5252    	 & C & 1.09 \\
NGC 5506 \#1  	 & C & 1.07 \\
NGC 5506 \#2  	 & C & 1.02 \\
NGC 5674	 &  C & 0.77\\
NGC 6300 	 &  C & 1.10 \\
ESO 103-G35 \#1 &  C & 0.95 \\
ESO 103-G35 \#2 & B & 0.61  \\
NGC 7172 \#1   	 & C & 0.88\\
NGC 7172 \#2   	 & C & 0.90 \\
NGC 7314   	 & C & 0.98 \\
NGC 7582        	 & C &  0.95 \\
NGC 7679         	 & C & 0.90 \\
\hline
\end{tabular}}
\caption{Summary of best fit results. Column 2: best fit model; column 3: reduced $\chi^2$ for the best fit model.}
\end{table}

\subsection{Photon Index}
\begin{figure}
\centerline{\resizebox{\hsize}{!}
{\includegraphics{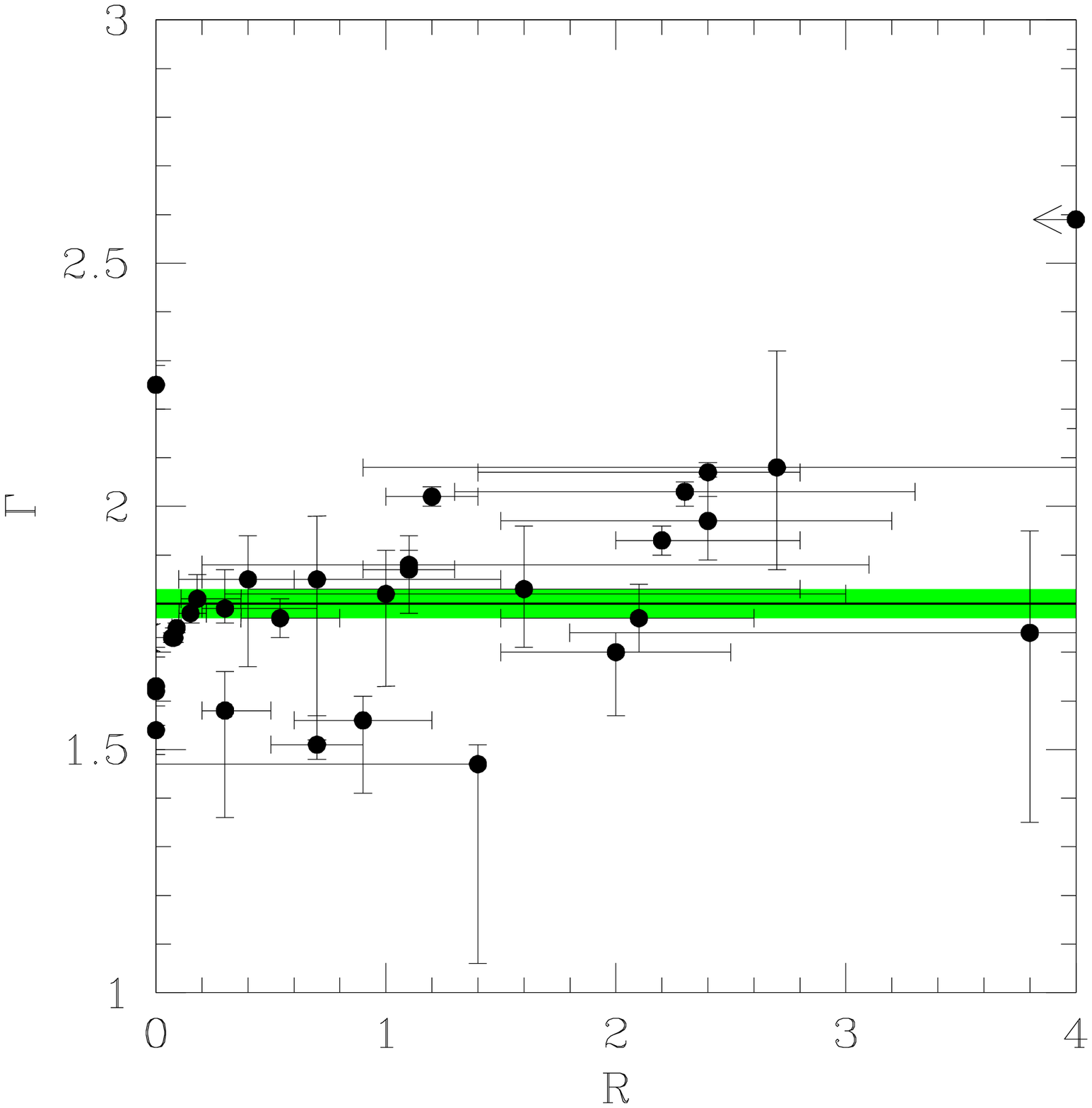}}}
\caption{\label{fig:rg}
\footnotesize{Photon index versus the ratio R between the normalizations of the reflected and transmitted components.
The line represents the average value of $\Gamma$ ($\Gamma_{\rm AV}=1.80\pm 0.03$).}}
\end{figure}

The wide energy range and the good statistics of our spectra allow a precise determination
of the photon index of the continuum power-law: 90\% errors are typically a few percent. The photon index
distribution is plotted in Fig. \ref{fig:rg} versus the ratio between the reflected and transmitted component (see below
for a discussion on this ratio).
 The data are from  the best fit model, either B or C, 
{\it according to the minimun reduced $\chi^2$}. The average photon index is
$\Gamma_{\rm AV}=1.79\pm0.01$. The 1$\sigma$ dispersion is 0.23. A similarly interesting parameter is the
``effective'' photon index, i.e. the slope of a simple power-law fitting the whole spectrum (model A). This
parameter is important for studies where the average X-ray SED of AGNs is approximated with a simple
power-law (for example, in most synthesis models of the cosmic background). The average photon index
derived from Table 3 is $\Gamma_{\rm E}=1.76\pm0.01$ with a 1$\sigma$ dispersion of 0.21. The small discrepancy
between the two evaluations of $\Gamma$ are due mainly to the contribution of the reflection component,
which is higher at high energies, so flattening the spectrum. The contribution of the high-energy cut-off
is small, since for most of the sources only a lower limit  can be determined (see below).
We note that our sample is the best currently available to determine the average $\Gamma$ of Seyfert
galaxies. 
The typical extent of this interval is from 2-3 keV up to 200 keV, and even in the
worst cases it is more than a decade (from 5-6 keV to 100 keV for NGC 4507). This is a significant
improvement with respect to past works, based on GINGA or ASCA data.
The photoelectric cut-off does not significantly reduce the energy range over which we observe the
intrinsic emission of the AGN and, on the other side, makes simpler the separation of the low energy
component.

\subsection{Reflection components}
{\bf Cold reflection:} the cold reflection component peaks around 30 keV. Therefore, the PDS instrument
allows a much better measurement of this feature than low energy instruments.
The model we used assumes a viewing angle of 30$^o$. Given the statistics in the PDS range, a variation
of this angle is analogous to a change in the normalization, since the spectral changes are too tiny
to be discriminated. The quantity used to measure the strength of reflection is the
ratio, R, between the normalizations of the reflected and transmitted components. A comparison between
Table 4 (model B, with no reflection) and Table 5 (model C, with reflection) illustrates the importance of a correct estimate
of the reflection component. In most cases adding a reflection component significantly improves the fit
(26 out of 31, Table 6). Moreover, in these cases a fit without reflection gives biased estimates
of the continuum components: for example, photon indexes and high energy cut-offs are systematically
lower in Table 4 than in Table 5.

Finally, it is worth noting that the errors associated to R are significantly
higher than those of the other free parameters. This is an effect of the
partial degeneracy between the different spectral components, in particular
the intrinsic power law, the high-energy cutoff and the cold reflection
component. If we freeze these parameters when estimating the errors on R,
we obtain much smaller errors.

\begin{figure}
\centerline{\resizebox{\hsize}{!}
{\includegraphics{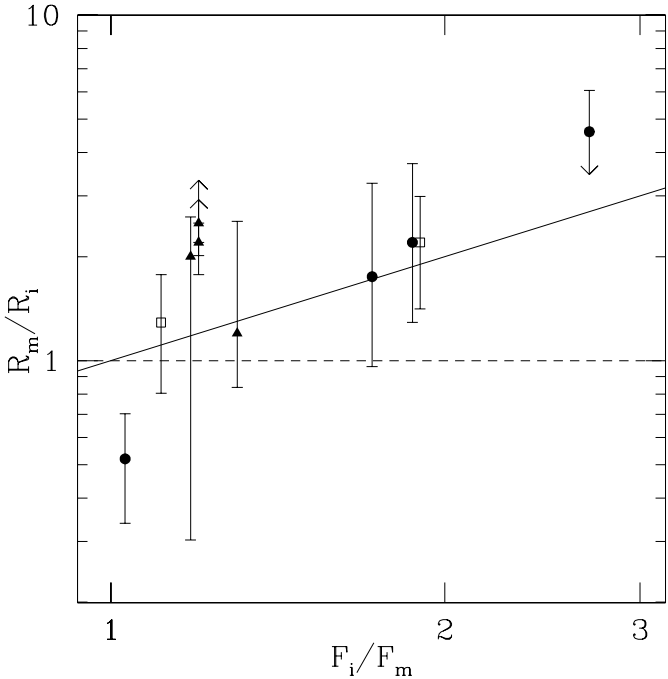}}}
\caption{\label{fig:fluxrefl2}
\footnotesize{Reflected component variations versus flux variations (see text for definitions)
The solid line is the correlation expected if the cold reflection component remains
constant when the intrinsic continuum varies. The dashed line is the correlation expected if
a variation in the intrinsic emission is immediately followed by a subsequent variation of the reflected component.
Sources with more than two observations are plotted as triangles (Cen A) and squares (NGC 4507).}}
\end{figure}

One of the main debates regarding the cold reflection component is whether it originates from
the accretion disk or from cold material farther from the center (eg. the putative ``torus'' of
unified models).
Our sample contains  multiple observations of 7 sources. This gives us the possibility
to study the long term variability of the cold reflection component, and thus to give an estimate
to the distance of the reflecting medium: if the reflector is the accretion
disk, we expect the reflection and transmitted component to be closely related.
In the other case, if the reflector distance from the center is greater than the light path
in the characteristic variability time of the primary component, we expect the reflected component
to remain constant. In order to study the correlation between the variability of the transmitted
and reflected components, we define for each pair of observations of the same source a
ratio f=F$_1$/F$_2$ where F is the 2-10 keV flux, and a ratio r=R$_2$/R$_1$.
If the reflected component varies together with the transmitted one, the quantities R$_i$
are expected to remain constant, and therefore r=1. If, on the other side, the absolute flux of the
reflected component remains constant, then R$_i  \propto$ 1/F$_i$, and r=f.
In Fig. \ref{fig:fluxrefl2} we plot the quantities r and f. For
 each pair, we chose the fluxes in order to have
F$_2 < F_1$ (then f$>1$). For the two sources with more than two observations
(NGC 4507, three observations, and Cen A, 5 observations), we have 2 and 4 pairs, respectively.
In these cases, F$_2$ is defined as the lowest measured flux among all the observations of the same source.
We also plot the two lines corresponding to r=1 and r=f.
The plot clearly shows that the r=f line better represents the measures. This is strong
evidence supporting a distant reflector, while the cold reflection from the disk is ruled out by the data.

If the cold reflection is mainly due to a distant reflector, the simplest structure of the circumnuclear
medium would be obtained with a single cold medium responsible for both the cold reflection
and absorption of the hard X-rays. However, in this scenario
the ratios between the reflected and transmitted components are somewhat surprising.
The model we used assumes a perfect efficiency reflection (i.e. no transmission) from an infinite
plane slab covering  a 2$\pi$ angle. The normalization factor of the model is relative to the intrinsic
continuum. Therefore, even assuming a perfectly face-on line of
sight, the ratio between the normalizations of the reflection model and of the power-law should be
not higher than 2. The value R=2 is for a completely Compton-thick reflector covering the whole
plane angle. Assuming a different geometry for the reflector, in particular a non-planar one, can
only slightly change the overall efficiency, as we can evaluate from the comparison of the PEXRAV
model with the one of Ghisellini et al. (1994), which assumes a toroidal geometry for the absorber:
the results are in agreement within a factor of $\sim 2$. The model of Ghisellini et al. 1994 also
shows that the reflection efficiency drops with the thickness of the reflector: at 30 keV the reflection
efficiency is 55\% for a Compton thick reflector
(N$_{\rm H} \sim 10^{25}$ cm$^{-2}$), 23\% for  N$_{\rm H} \sim 10^{24}$  cm$^{-2}$
and only 8\% for   N$_{\rm H} \sim 10^{23}$  cm$^{-2}$.
The values for the ratio we obtained (Table 5) are in many cases too high to be explained assuming
a homogeneous reflector having the same column density as measured in absorption.
In the extreme case of  NGC 1365 the best fit value, R=3.8, is not acceptable. We discuss
this case in the Appendix. Out of  the other 20 objects, 17 have the value of R higher than the
maximum allowed by the models described above, the only  exceptions being NGC 526a,
NGC 2110, NGC 4258 and Cen A. Part of the discrepancy can be due, in some cases, to the fact that the
source is observed in a low state, and therefore the value of R is high, as explained above. However,
this cannot be the case for 18 sources out of 21. Moreover, even in those cases where multiple
observations are available, R is too high even in the high flux states.
The only possible explanation is that
the absorber is not
homogeneous, and a Compton thick medium covers a significant fraction of the solid angle (but
not the line of sight).  We note that other recent studies on variability of N$_{\rm H}$ support the view of
a non homogeneous absorber (Risaliti et al. 2002).

Finally, our sample is useful to investigate the putative correlation between the photon index and the ratio
R. Such correlation has been first claimed by Zdziarski et al. 1999, and further analyzed by Matt (2000) and
Petrucci et al. (2001), taking advantage of the broad BeppoSAX band. At present it is still debated,
since the two parameters are not independent in the fitting procedure, an increase of $\Gamma$ being
compensated by a decrease of R. In Fig. \ref{fig:rg} we plot the two quantities.
A slight increase of $\Gamma$ for R$>2$ is present, but the correlation is weaker than that found by
Petrucci et al. 2001 for a sample of bright Seyfert 1s. We can explain this discrepancy using an argument
of Matt 2000. In order to give a warning about the reality of this correlation, this author shows
that if the soft component of the Seyfert 1 MKN 841 is fitted with
a warm absorber instead of a blackbody, the best fit photon index changes by $\sim 0.4$ and, as a consequence,
R changes by a factor or $\sim 2$. This example shows how strong is the dependence among the different
components of the continuum models. As we already pointed out above, our fits are made safer
by the presence of the photoelectric cut-off, which clearly separates the soft component from the others.
As a consequence, the statistical dependence among the two quantities is weaker in our sample.
Therefore, the results plotted in Fig. \ref{fig:rg} suggest that the correlation is probably not real.

{\bf Warm reflection:} Adding a warm reflection component improves the fit only for three sources (IRAS 05189-2524,
NGC 4388 and Cen A, see Table 5). In these cases the reflection efficiency is between 1\% and 6\%. In all the other
cases we were able to estimate an upper limit for the warm reflection of the order of a few percent of the intrinsic
component.

\subsection{The high-energy cut-off}
\begin{figure}
\centerline{\resizebox{\hsize}{!}
{\includegraphics{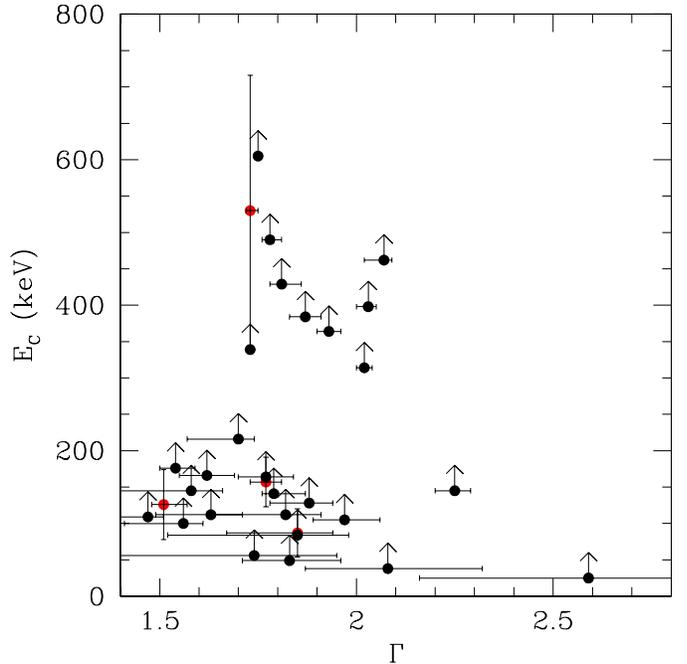}}}
\caption{\label{fig:gc}
\footnotesize{High energy cut-off versus photon index of the primary power law.}}
\end{figure}

The energy of the exponential cut-off is plotted versus the photon index in Fig. \ref{fig:gc}.
 In most cases we have only
lower limits, however some interesting conclusions can nevertheless be drawn. Many lower limits are too
low to give significant constraints. However, we note a significant fraction of detections and lower limits
higher than 300 keV: 6 sources out of 20 have have E$_{\rm C}>$300 keV. We conclude that a photoelectric
cut-off in the 100-300 keV band is not an ubiquitous property of Seyfert galaxies.

\subsection{The iron line}
\begin{figure}
\centerline{\resizebox{\hsize}{!}
{\includegraphics{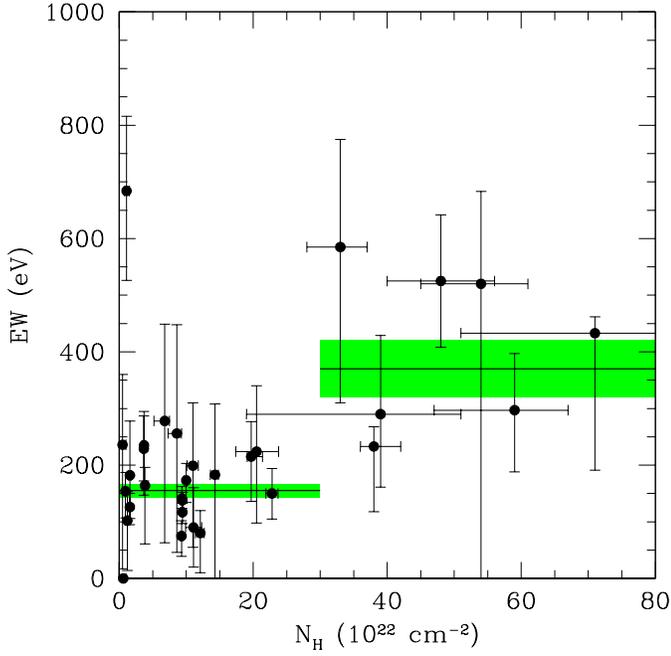}}}
\caption{\label{fig:eqnh}
\footnotesize{Observed equivalent width of the iron K$\alpha$ line versus the absorbing column density.
The two lines are the average equivalent widths for the subsamples with N$_{\rm H} < 3\times 10^{23}$ cm$^{-2}$ and
N$_{\rm H} > 3\times 10^{23}$ cm$^{-2}$, respectively. The shaded regions are delimited by the 1 $\sigma$
errors on the
averages (for the low-$N_{\rm H}$ objects we excluded the point in the upper-left of the plot from the calculation.}}
\end{figure}

{\bf Equivalent width:}
We detect an iron K$\alpha$ line in all the observations but one (NGC 7679). The equivalent width, EW,
is in the range 100-300 eV for sources with low absorbing column density (N$_{\rm H} < 10^{23}$ cm$^{-2}$),
in agreement with previous studies, and  with the values found in Seyfert 1s (Gilli et al. 1999). In sources
with a larger N$_{\rm H}$, the continuum at the energy of the line is partially absorbed, and the EW of the
iron line is therefore expected to be higher.
In Fig. \ref{fig:eqnh}, where we plot the EW of the iron line versus
N$_{\rm H}$.  A correlation is present: sources with N$_{\rm H} > 3\times 10^{23}$ cm$^{-2}$ have on average
an EW higher that sources with N$_{\rm H} < 3\times 10^{23}$, even if the errors are too large to draw solid
conclusions.

\begin{figure}
\centerline{\resizebox{\hsize}{!}
{\includegraphics{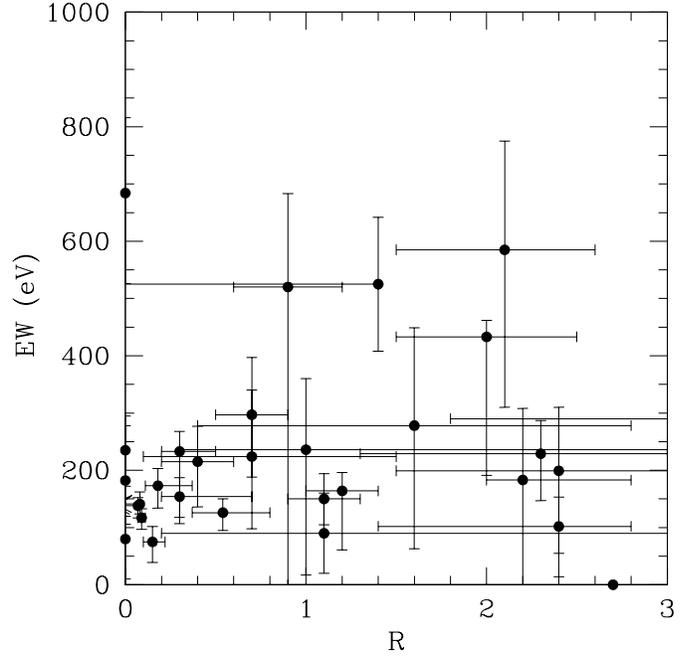}}}
\caption{\label{fig:eqr}
\footnotesize{Equivalent width of the iron K$\alpha$ line versus the cold reflection component (normalized
to the primary power law).}}
\end{figure}

This correlation suggests that the iron line is produced, at least in part, by a reflector different
from the accretion disk. Otherwise, we would not expect any increase of EW with N$_{\rm H}$, since the
line would be as much absorbed as  the continuum component.
In Fig. \ref{fig:eqr} we plot the line EW versus the cold reflection ratio, R. A correlation is non clear, even if the errors
are again too large to rule out an average increase of EW with R. We also note that in this kind of correlations,
where the expected variation of EW is a factor of no more than 2-3, an important role could be played
be the inclination of the accretion disk with respect of the line of sight: the primary continuum emission from
an edge-on disk is weaker than that of a face-on disk with the same intrinsic luminosity. Therefore,
if the iron line is partly produced by an outer reflector, the EW depends on the inclination angle. The effect
is a further dilution of the correlations we are trying to investigate.

{\bf Line energy and width:} In all cases the measured peak energy is compatible with neutral or little
ionized iron. In 9 cases out of 31, we measure a line width significantly larger than 0. For all these
cases we analyzed the model residuals to look for asymmetries in the line profile. We did not find
any significant asymmetry, except than in one case, NGC 6300, which is discussed in the Appendix, and
represents a convincing example of a relativistic line in Seyfert 2s.

\subsection{Soft excess}

In this work the soft emission is not analyzed in detail. A single thermal component,
plus the warm reflection component, are used to fit the
points at energies lower than the photoelectric cut-off. This gives on average good fits. The residuals
at low energies are in no case high enough to alter the best fit values of the other high energy parameters
which we are more interested to. In the simple model A we fit the soft component with a power-law
with a free photon index.
The ratio between the absorbed and the unabsorbed power-laws, reported in Column 6 of Table 3, shows that
the normalization at 1 keV of the ``soft'' component is typically a few percent of that of the absorbed
component. This confirms the separation of our fits in two almost independent parts: the high energy components
depends only on the data at energies higher than the cut-off, while the soft component is entirely determined
by the low energy counts.

\section{Conclusions}

We presented a homogeneous analysis of 31 spectra of 20 Compton thin Seyfert 2s observed with
the BeppoSAX satellite.
We fitted the spectra with multi-component models, including a cold and warm reflection,
an iron line, a low-energy thermal emission, and a primary continuum modeled with an absorbed power law
with a high-energy cut-off.
We also investigated the main correlations among the best fit parameters, in order to test the unified models.
The main conclusions are the following.
\begin{enumerate}
\item Testing Unified models:
The spectra of the objects in our sample are in good agreement with the standard unified model of AGNs:
the primary continuum is well represented by a power law with photon index $\Gamma=1.79\pm0.01$, with a dispersion
$\sigma(\Gamma)=0.23$. This is in agreement with previous studies on the hard X-ray emission of Seyfert 2s
(Turner et al. 1997, Nandra \& Pounds 1989, Smith \& Done 1997, Turner \& Pounds 1989), and with the
hard X-ray spectra of Seyfert 1s. All the objects but one show significant cold absorption by gas with
column density $N_{\rm H} >10^{22}$ cm$^{-2}$, in agreement with the optical classification of type 2 or 1.8/1.9.
The only exception, NGC 7679,  is an intriguing object, with no significant absorption in excess to Galactic
in the X-rays, but also with no evidence of broad lines in the optical (see the Appendix and Della Ceca et al. 2001
for further discussion).
We also fitted the hard spectrum of our objects with a simple absorbed power law, in order to compute
the average observed photon index, $\Gamma_{\rm eff}$. We found  $\Gamma_{\rm eff}=1.76\pm 0.01$, with a dispersion
$\sigma(\Gamma_{\rm eff})=0.21$.

An iron K$\alpha$ emission line at energies $\sim 6.4-6.6$ keV is found in all the objects but one (NGC 7679).
In agreement with the unified model, the average equivalent width of the line is higher in objects with higher absorbing
column density. In 6 cases we found a broad profile of the line. In one case (NGC 6300) this profile is best fitted with a
relativistic model. In the other 5 cases, the statistics is not enough to test models of relativistic lines.

\item A reflected component is present in most of the objects (17 out of 21).
A debated issue on X-ray spectra of Seyferts is whether the reflection is due to the accretion disk or to a farther
reflector.
Comparing the variations of the direct and
reflected components in the 7 sources with multiple observations, we find that the variability data are better reproduced
assuming that the reflector is located far ($>1$ light year) from the primary source.

\item The ratio between the reflected and direct components is on average too high for a homogeneous reflector with
the same column density as measured in absorption in our spectra. This is an indication that the circumnuclear medium
is not homogeneous and probably a Compton thick gas cover a significant fraction of the solid angle.
This structure is also suggested by X-ray variability studies of Seyfert galaxies (Risaliti et al. 2002) and will be further
tested in a forthcoming detailed study of variability of this sample of AGNs (Risaliti et al. 2002, in preparation).

\end{enumerate}

\acknowledgements
I am grateful to the referee, Dr. K. Iwasawa, for helpful comments. 
This work was partially supported by the Italian
Ministry for University and Research (MURST) under grant Cofin00-02-36.

\appendix
\section{Analysis of single sources}

\noindent {\bf NGC 526a:}
The same set of BeppoSAX data on this source have been analyzed by Landi
et al. (2001). The results obtained by these authors are in agreement with
ours, within the errors. In particular, it is confirmed the relatively flat
continuum spectrum: the best fit values for the photon index of the intrinsic
power law are in the 1.47-1.6 range, in agreement with our result,
$\Gamma=1.54^{+0.05}_{-0.04}$. 

\noindent {\bf NGC 1365:}
NGC 1365 is one of the few sources for which our models do not provide a good fit
($\chi^2_{\rm r}$=1.26 in model C).
A more careful analysis of this object was performed by
Risaliti et al. (2000). According to this work, the PDS data of NGC 1365 could be
contaminated by the nearby active galaxy NGC 1386. Moreover, NGC 1365 shows an high
and complex variability during the BeppoSAX observation. However, the best fit
values in Risaliti et al. 2000 are in agreement with those obtained with our model C.

\noindent {\bf NGC 2992:}
This source shows the strongest long-term variability in our sample. Our analysis is
in agreement with that of Gilli et al. (2000).

\noindent {\bf IRAS 05189-2524:}
The BeppoSAX observation of this source was analyzed in detail by. Severgnini et al. 2000
The reflected component is not required from the fit ($\Delta \chi^2=2$ with one more
free parameter). The fit obtained with model B is in agreement with that of  Severgnini et al. (2000).
The poor statistics ($\chi^2_{\rm r}$=1.2) is mainly due to an excess in the PDS data around 20 keV.
The statistics is however not good enough to understand whether this excess is real or not.
We note that given the high noise in the PDS data, Severgnini et al. 2000 chose not to use
these data in their analysis.

\noindent {\bf NGC 4258:}
NGC 4258 is a nearby galaxy (estimated distance $\sim 7$ Mpc) hosting a weak AGN.
In this object the integrated X-ray emission of galactic sources is not negligible with respect
to the AGN contribution. Therefore, in models B and C we added an extra bremmshtralung component
to take into account for the galactic contribution. The best fit temperature (kT=7 keV) and
normalization are in agreement with those estimated in 5 different measures performed with
ASCA (Reynolds et al. 2000). Our results is in agreement with the more detailed analysis of
the same data performed by Fiore et al. (2001).

\begin{figure}
\centerline{\resizebox{\hsize}{!}
{\includegraphics{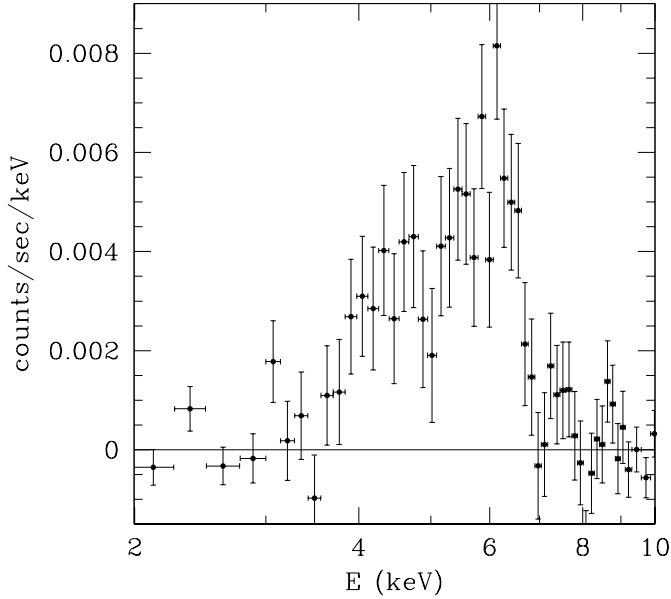}}}
\caption{\label{fig:n6300}
\footnotesize{Residuals of the best fit model for NGC 6300, after the removal of the relativistic
line from the model.}}
\end{figure}

\noindent {\bf NGC 6300:}
In Fig. \ref{fig:n6300} we plot the residuals of the best fit of NGC 6300, obtained with model C, but without the
Gaussian component that fits the iron K$\alpha$ line. The asymmetry of the line is apparent:
a clear ``red wing'' extends down to $\sim 4$ keV, while a broad blue wing is absent.
We fitted this excess with a relativistic line model. We used the DISKLINE model in XSPEC,
leaving all the parameters free. The best fit is obtained with a strongly relativistic profile: the
best fit inner radius of the emitting region is 6.5 Shwarzchild radii; the inclination is $\sim 30^o$.
The improvement of the fit, with respect to the simple gaussian fit, is $\Delta \chi^2 =27$.
We also tried a diskline modelem for a rotating disk, but the fit is worse, since this model
can well reproduce the broad red wing, but not the narrow peak at E=6.1 keV.
We note that a possible warning on this result could come from the high
N$_{\rm H}$
of this source (N$_{\rm H} \sim 2\times 10^{23}$ cm$^{-2}$). This implies that the continuum at the
low energy peak of the line is partially absorbed. This could introduce systematic errors, making the reality
of the red wing of the line less certain than it appears in our fit.

Another possible scenario is the one proposed by Guainazzi (2001): a good fit
of the MECS spectrum can be obtained accepting very high values of the ratio R
between the reflected and intrinsic component (R$\sim 4$). In this way the flat profile of
the cold reflection spectrum can reproduce the excess between 4 and 6 keV
which has been interpreted above as a relativistic wing of the iron line.
Such high values of R are possible, as suggested by Guainazzi (2001), if the
intrinsic flux is variable, since the observed reflection component is
produced by 
radiation emitted at a different time with respect to the observed
primary emission.

\noindent {\bf ESO 103-G35:}
The BeppoSAX data of this source have been analyzed by Wilkes et al. (2000). Even if the models
used in our work are slightly different, the basic results are in agreement with these authors.
The only noticeable discrepancy  is about the high energy cut-off measurements. In the first
observation we find E$_{\rm C} = 84^{+134}_{-33}$ keV. This is the lowest measured
cut-off energy in our sample,
but it is significantly higher than that found by Wilkes et al. 2000 (E$_{\rm C} =29\pm 10$ keV).

\noindent {\bf NGC 7172:}
There are two published works on the analysis of the same data (Akylas et al. 2001 and Dadina et al. 2001).
Akylas et al. 2001 propose a model with no reflection. Their best fit parameters are
in agreement with our model B  (Table 4). Dadina et al. 2001 find that the inclusion of a cold reflection
slightly increase the goodness of the fit. The best fit slope of the continuum is significantly higher in this model
Our results are in agreement with  Dadina et al. 2001.
Our best fit model C (Table 5) is only slightly better than model B, and predicts a steeper intrinsic continuum
($\Gamma\sim 1.9$), than model B ($\Gamma\sim 1.7$, in agreement with Akylas et al. 2001).

\noindent {\bf NGC 7582:}
Turner et al. (2000) present a detailed analysis of the same BeppoSAX data. The results are compatible,
but these authors prefer a different interpretation for the high energy emission in excess of the simple
powerlaw, which we fit with a strong cold reflection component. Arguing that the required reflection component
is too high, they suggest that the intrinsic continuum (which is completely covered by gas with N$_{\rm H} \sim 1.4 \times
10^{23}$ cm$^{-2}$, in agreement with our results), is also absorbed by a second screen with
N$_{\rm H} \sim 1.6 \times 10^{24}$ cm$^{-2}$, and a covering factor of $\sim 60$\%.

\noindent {\bf NGC 7679:}
This source is peculiar within our sample, for it is the only one with negligible X-ray absorption.
The optical classification (pure Seyfert 2) suggests that the BLR emission is obscured by dust by
at least 3-4 magnitudes . This, assuming a galactic dust-to-gas ratio, would imply an absorbing
column density of at least 6-7 10$^{21}$ cm$^{-2}$. We can explain this discrepancy in two ways:
a) the dust-to-gas ratio could be much higher than galactic, or the dust could
be associated to a warm absorber; b) there could be a ``hole'' in the absorber of the nuclear source,
so that the X-ray emission from the accretion disk is unabsorbed, but the broad lines, emitted by
the larger Broad line Region, are obscured. This scenario is not unplausible, given the recent evidence
of a complex structure in the Seyfert 2 absorbers (Risaliti et al 2002).
Both the fits with and without the reflected component are acceptable (see Table 4 and 5).
In our analysis the best reduced $\chi^2$ is obtained with a steep $\Gamma\sim2.1$ power law plus a
strong reflected component. However, in a recent study of the same data, Della Ceca et al. (2001) propose
a best fit model in agreement with the one in Table 4, without reflection, and with a photon index
$\Gamma \sim 1.75$.

\clearpage

\begin{figure*}
\centerline{\resizebox{\hsize}{!}
{\includegraphics{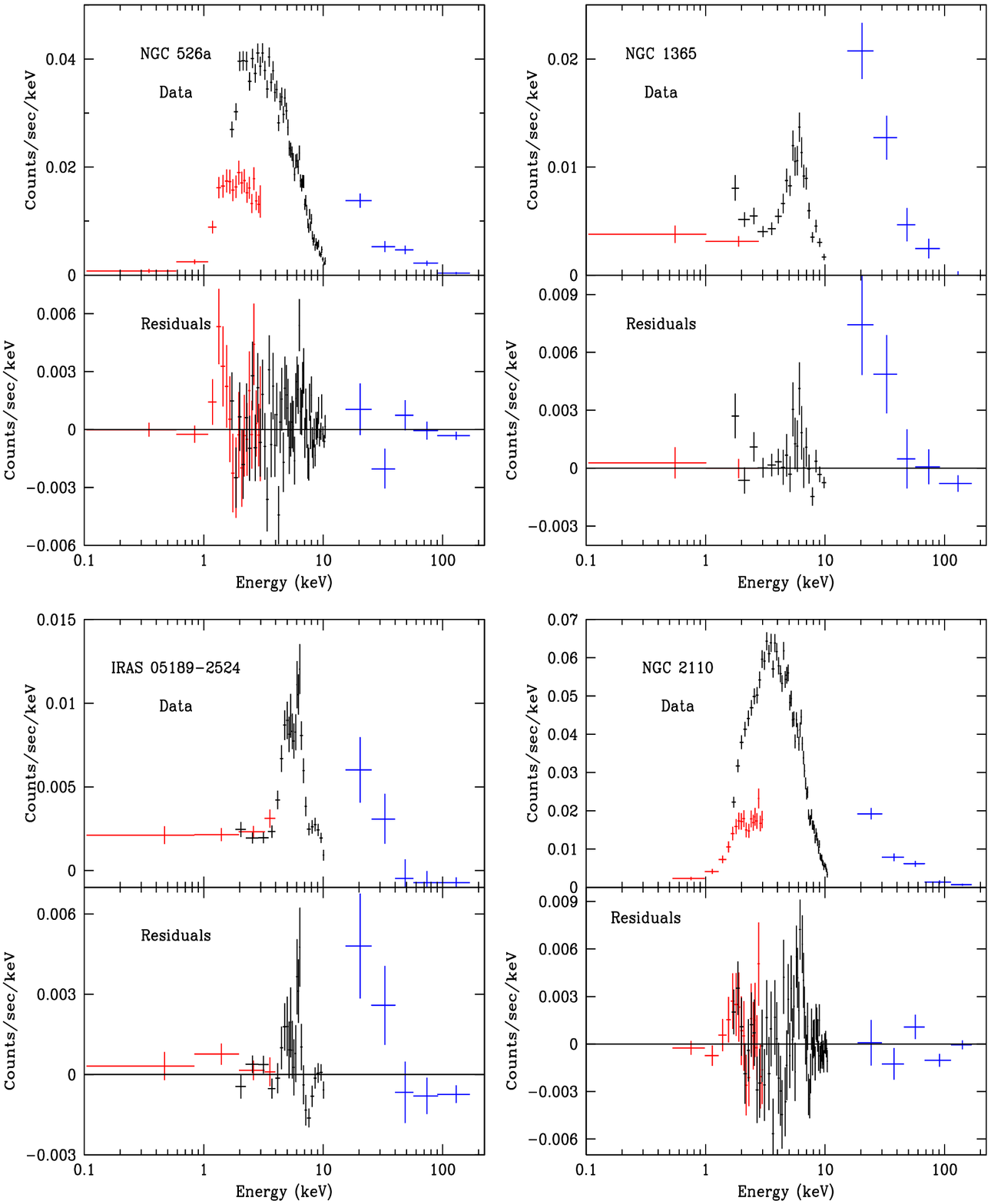}}}
\caption{\label{fig:s1}
\footnotesize{Spectra and residuals for the whole sample. Residuals are
calculated using model A (absorbed power law plus a second power law to fit the
soft component). }}
\end{figure*}
\addtocounter{figure}{-1}

\begin{figure*}
\centerline{\resizebox{\hsize}{!}
{\includegraphics{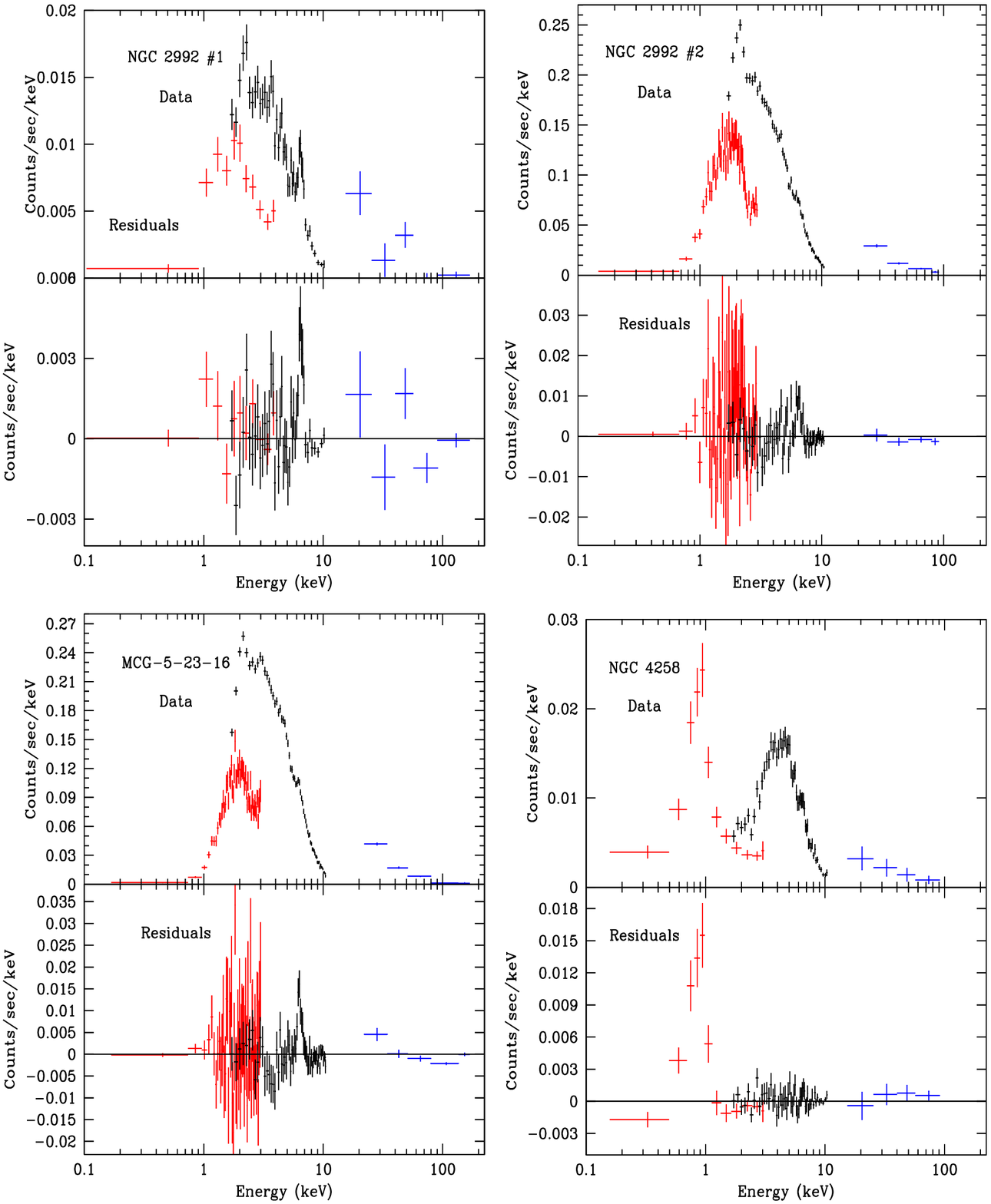}}}
\caption{\label{fig:s2}
\footnotesize{Continued.}}
\end{figure*}
\addtocounter{figure}{-1}

\begin{figure*}
\centerline{\resizebox{\hsize}{!}
{\includegraphics{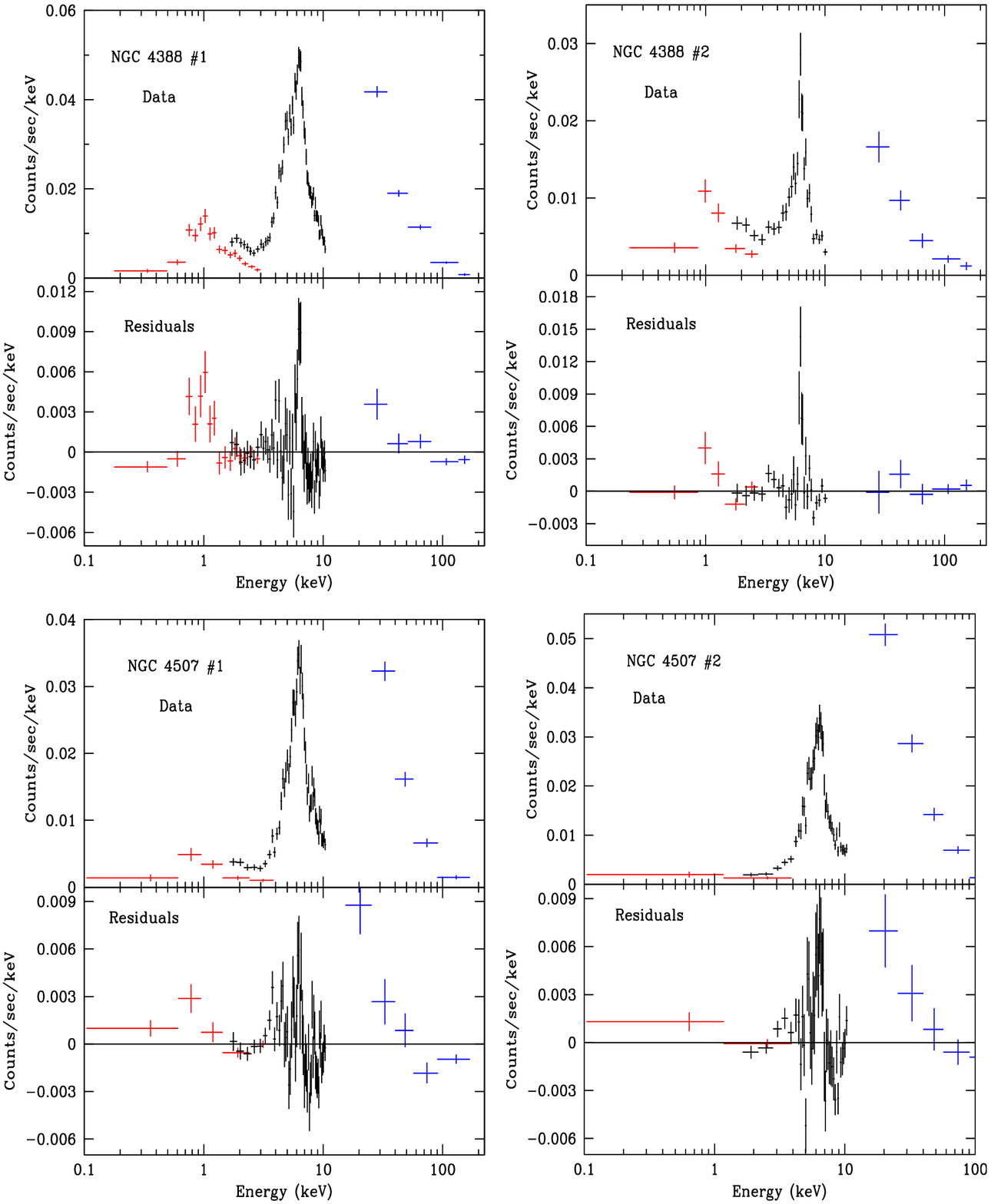}}}
\caption{\label{fig:s3}
\footnotesize{Continued.}}
\end{figure*}
\addtocounter{figure}{-1}

\begin{figure*}
\centerline{\resizebox{\hsize}{!}
{\includegraphics{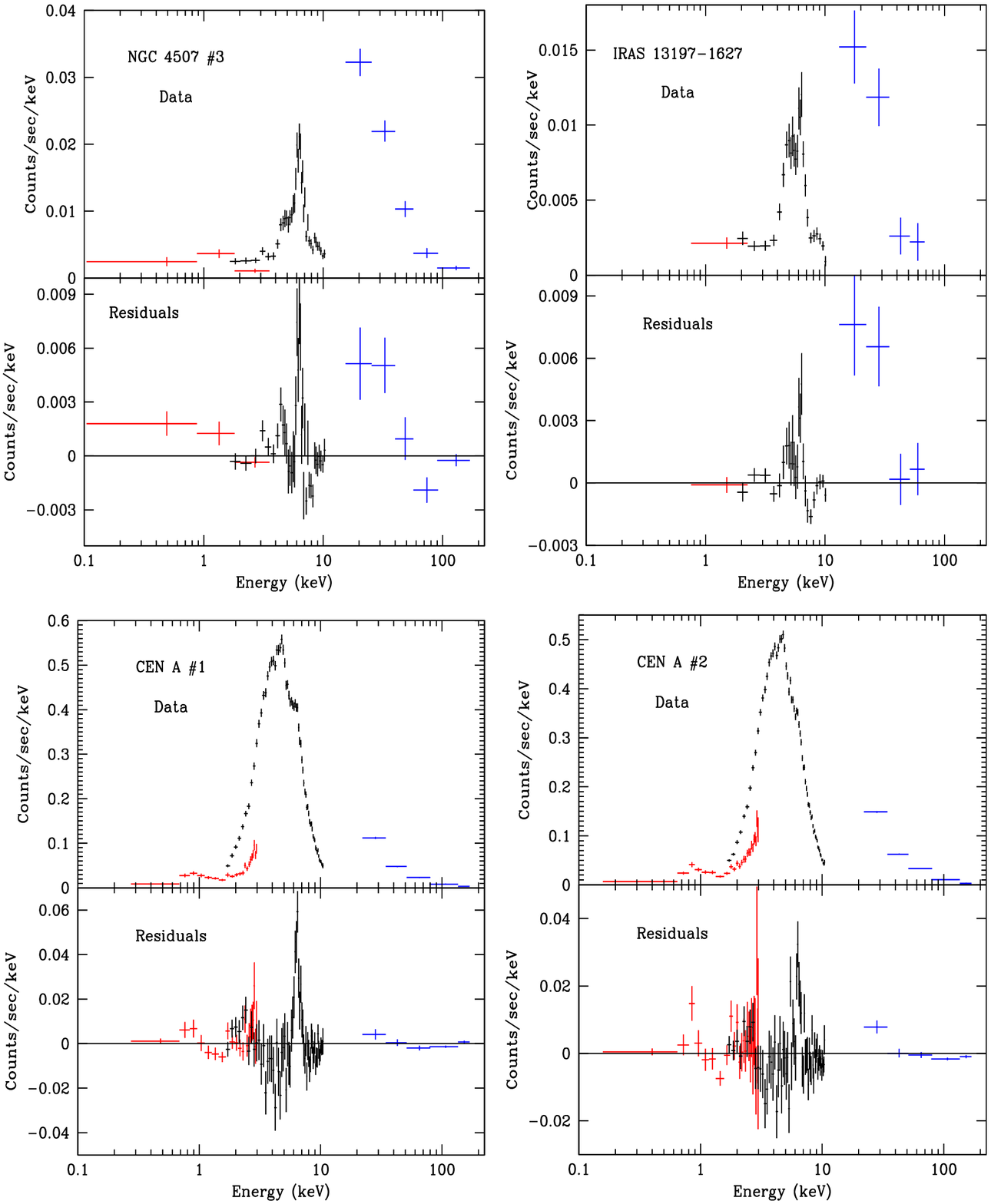}}}
\caption{\label{fig:s4}
\footnotesize{Continued.}}
\end{figure*}
\addtocounter{figure}{-1}

\begin{figure*}
\centerline{\resizebox{\hsize}{!}
{\includegraphics{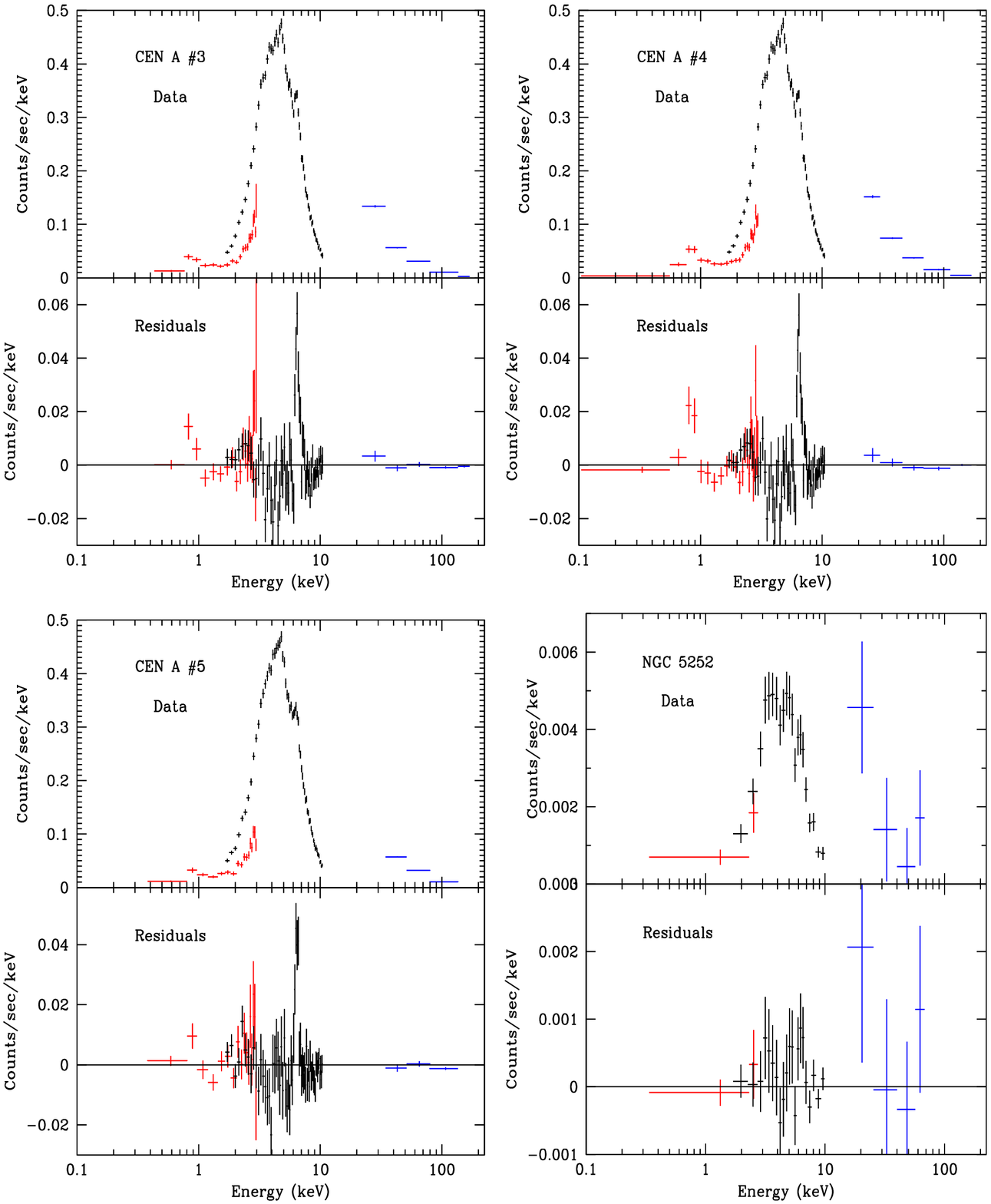}}}
\caption{\label{fig:s5}
\footnotesize{Continued.}}
\end{figure*}
\addtocounter{figure}{-1}

\begin{figure*}
\centerline{\resizebox{\hsize}{!}
{\includegraphics{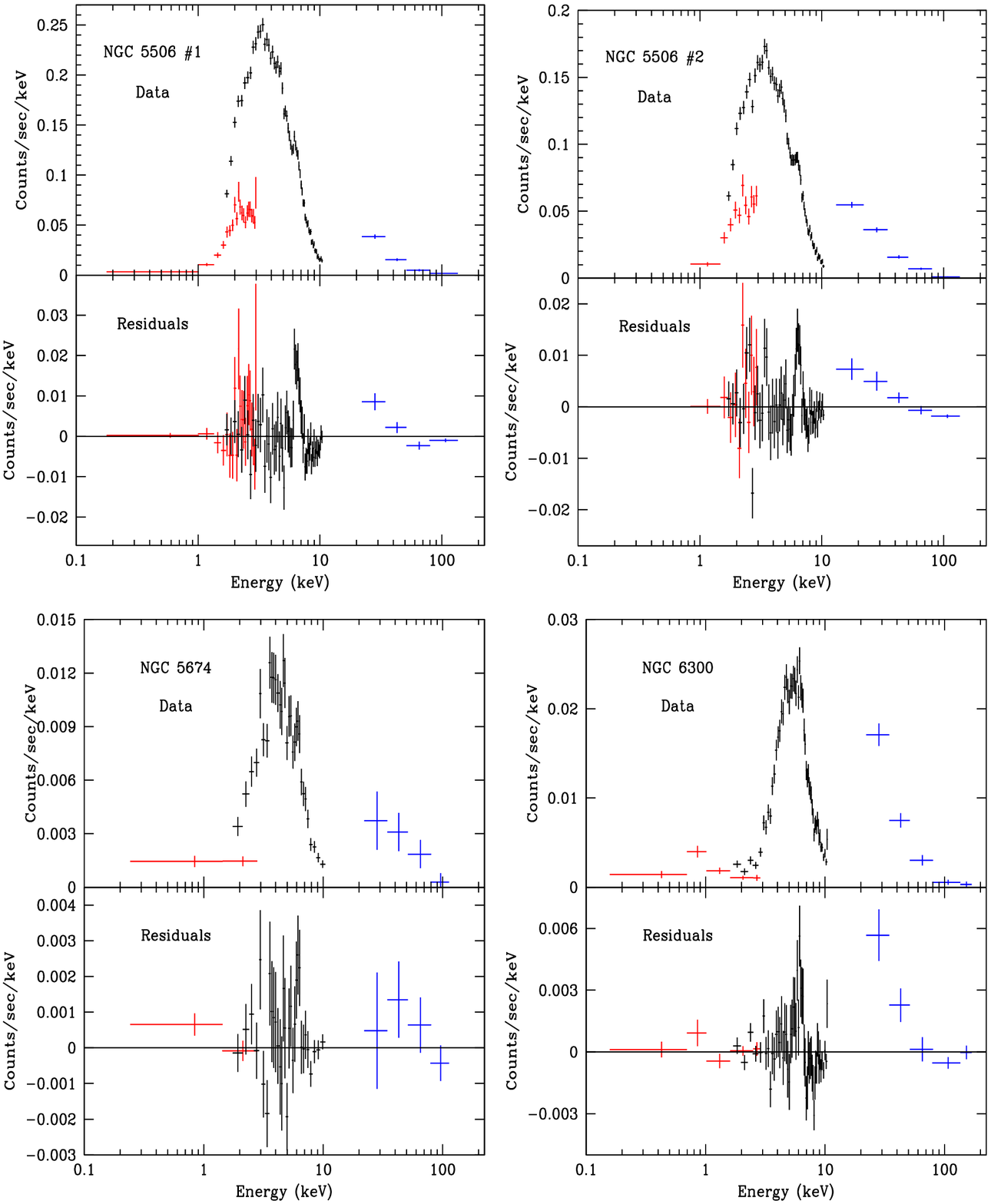}}}
\caption{\label{fig:s6}
\footnotesize{Continued.}}
\end{figure*}
\addtocounter{figure}{-1}

\begin{figure*}
\centerline{\resizebox{\hsize}{!}
{\includegraphics{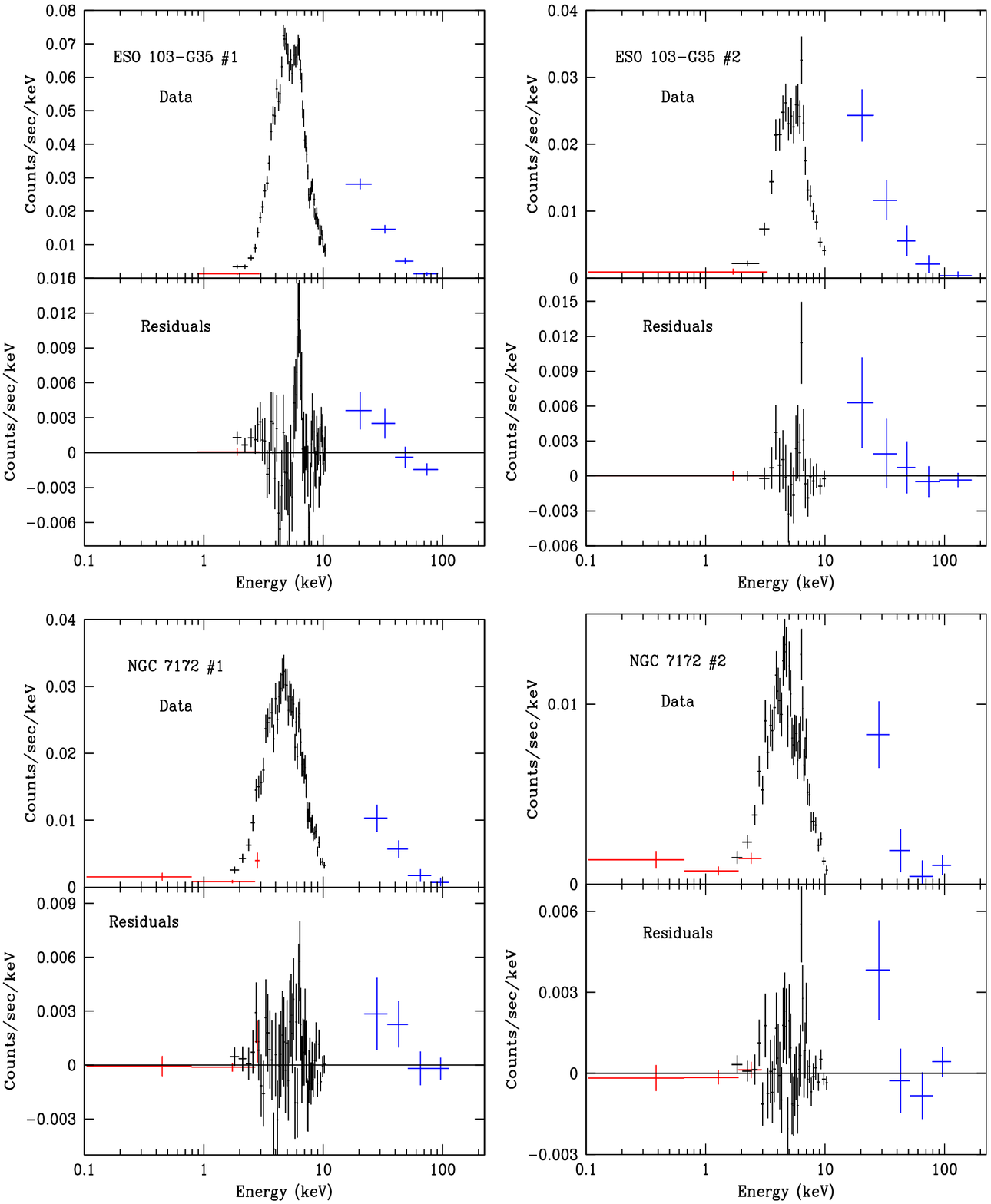}}}
\caption{\label{fig:s7}
\footnotesize{Continued.}}
\end{figure*}
\addtocounter{figure}{-1}

\begin{figure*}
\centerline{\resizebox{\hsize}{!}
{\includegraphics{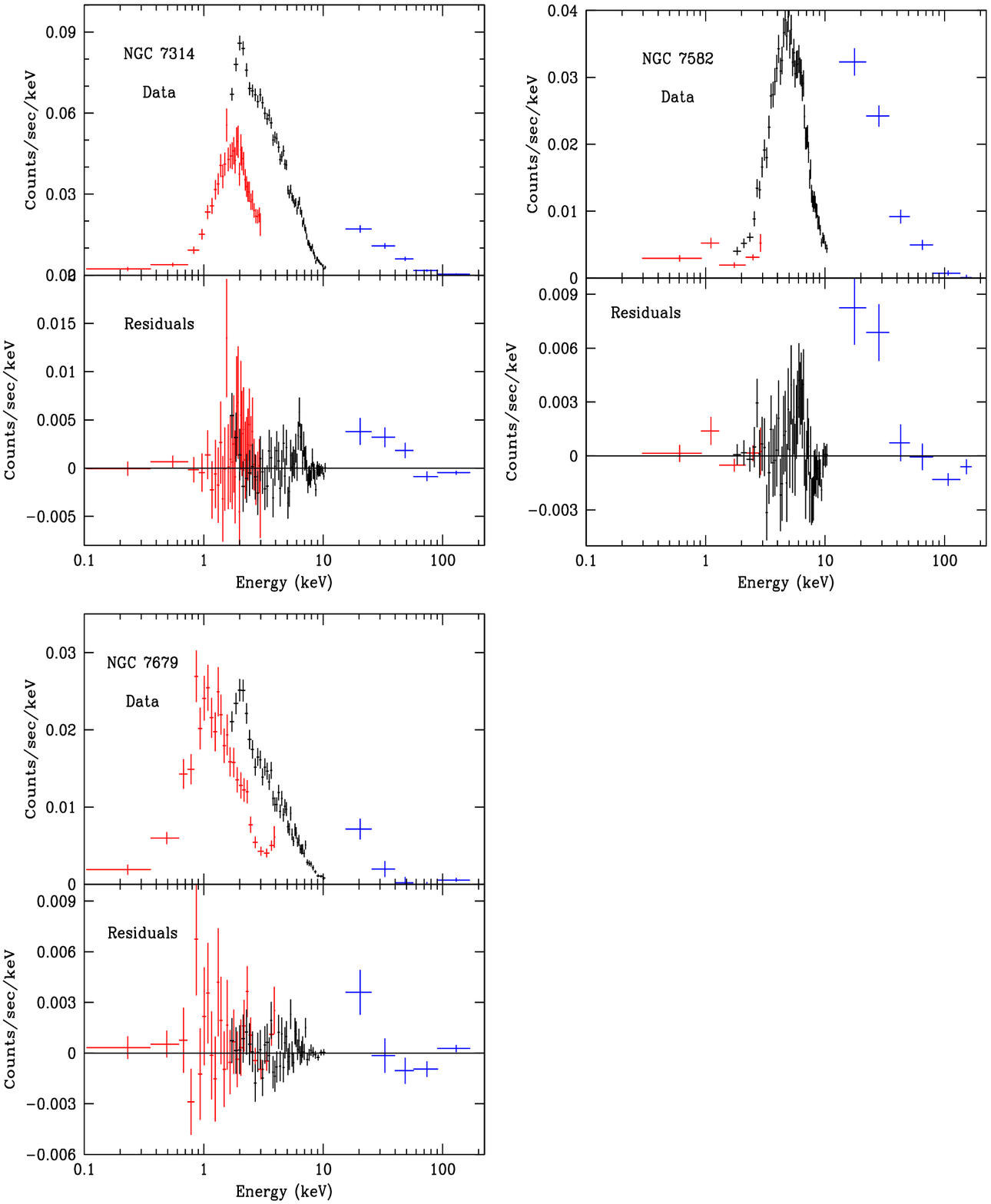}}}
\caption{\label{fig:s8}
\footnotesize{Continued.}}
\end{figure*}

\end{document}